\pdfoutput=1

\documentclass[aps, pra, reprint, floatfix, superscriptaddress]{revtex4-2}
\usepackage{graphicx}   
\usepackage{siunitx}
\usepackage{bm}
\usepackage{physics}
\usepackage{dcolumn}
\begin{document}

\title{Layer-dependent optically-induced spin polarization in InSe}

\author{Jovan Nelson}
\affiliation{Applied Physics Program, Northwestern University, Evanston, Illinois 60208, USA}

\author{Teodor K. Stanev}%
\affiliation{Department of Physics and Astronomy, Northwestern University, Evanston, Illinois 60208, USA
}%

\author{Dmitry Lebedev}
\affiliation{Department of Material Science $\&$ Engineering, Northwestern University, Evanston, Illinois 60208, USA
}%

\author{Trevor LaMountain}
\affiliation{Applied Physics Program, Northwestern University, Evanston, Illinois 60208, USA
}%

\author{J. Tyler Gish}
\affiliation{Department of Material Science $\&$ Engineering, Northwestern University, Evanston, Illinois 60208, USA
}%

\author{Hongfei Zeng}
\affiliation{Department of Physics and Astronomy, Northwestern University, Evanston, Illinois 60208, USA
}%

\author{Hyeondeok Shin}
\affiliation{Computational Science Division, Argonne National Laboratory, Lemont, Illinois 60439, USA}

\author{Olle Heinonen}
\affiliation{Material Science Division, Argonne National Laboratory, Lemont, Illinois 60439, USA}
\affiliation{Present and permanent address: Seagate Technology, 7801 Computer Ave, Bloomington, Minnesota 55435, USA}

\author{Kenji Watanabe}
\affiliation{Research Center for Functional Materials, National Institute for Materials Science, 1-1 Namiki, Tsukuba 305-0044, Japan}

\author{Takashi Taniguchi}
\affiliation{International Center for Materials Nanoarchitectonics, National Institute for Materials Science, 1-1 Namiki, Tsukuba 305-0044, Japan}

\author{Mark C. Hersam}
\affiliation{Applied Physics Program, Northwestern University, Evanston, Illinois 60208, USA}

\affiliation{Department of Material Science $\&$ Engineering, Northwestern University, Evanston, Illinois 60208, USA}
\affiliation{Department of Chemistry, Northwestern University, Evanston, Illinois 60208, USA}
\affiliation{Department of Electrical Engineering and Computer Science, Northwestern University,
Evanston, Illinois 60208, USA}

\author{Nathaniel P. Stern}
\email{n-stern@northwestern.edu}
\affiliation{Applied Physics Program, Northwestern University, Evanston, Illinois 60208, USA}

\affiliation{Department of Physics and Astronomy, Northwestern University, Evanston, Illinois 60208, USA
}%

\date{\today}

\begin{abstract}

Optical control of spin in semiconductors has been pioneered using nanostructures of III-V and II-VI semiconductors, but the emergence of two-dimensional van der Waals materials offers an alternative low-dimensional platform for spintronic phenomena.
Indium selenide (InSe), a group-III monochalcogenide van der Waals material, has shown promise for opto-electronics due to its high electron mobility, tunable direct bandgap, and quantum transport. In addition to these confirmed properties, there are  predictions of spin-dependent optical selection rules suggesting potential for all-optical excitation and control of spin in a two-dimensional layered material.
Despite these predictions, layer-dependent optical spin phenomena in InSe have yet to be explored. Here, we present measurements of layer-dependent optical spin dynamics in few-layer and bulk InSe. Polarized photoluminescence reveals layer-dependent optical orientation of spin, thereby demonstrating the optical selection rules in few-layer InSe. Spin dynamics are also studied in many-layer InSe using time-resolved Kerr rotation spectroscopy. By applying out-of-plane and in-plane static magnetic fields for polarized emission measurements and Kerr measurements, respectively, the $g$-factor for InSe was extracted. Further investigations are done by calculating precession values using a $\textbf{k} \cdot \textbf{p}$ model, which is supported by \textit{ab-initio} density functional theory. Comparison of predicted precession rates with experimental measurements highlights the importance of excitonic effects in InSe for understanding spin dynamics. Optical orientation of spin is an important prerequisite for opto-spintronic phenomena and devices, and these first demonstrations of layer-dependent optical excitation of spins in InSe lay the foundation for combining layer-dependent spin properties with advantageous electronic properties found in this material.

\end{abstract}


\maketitle

\section{Introduction}

The growing demand for rapid computation and high-density storage has driven the search for dynamic and fast control over spin in solid-state materials. Traditional manipulation of spin with static magnets can be slow, but all-optical spin orientation and manipulation, enabled by polarization-dependent optical selection rules in a material, offer opportunities for high-speed, non-invasive, and magnet-free control over spin information~\cite{Zutic2004}. To this end, a key ingredient of spin injection is optically-induced spin orientation (OISO), which has been exploited in III-V and II-VI semiconductors for spintronic applications such as spin transport~\cite{Crooker2007}, spin memory~\cite{Kroutvar2004}, and spin coherence~\cite{Greilich2006}.
More recently, optical orientation and control of spin have been explored in atomically-thin materials. These systems bring new possibilities such as layer-by-layer engineering in two-dimensional (2D) heterostructures. This approach has implications for spin physics, allowing the paring of materials with complimentary physical properties such as large spin-orbit coupling for valley-spin manipulation (WSe$_2$) and high conductance for electronics (graphene)~\cite{Geim2013}. In particular, a large proximity effect in WSe$_2$/Graphene heterostructures has been demonstrated, presenting unique control over valley-spin dynamics~\cite{LiGr}. Group-VI transition metal dichalcogenides (TMDs) are the canonical examples of 2D materials with non-trivial optical and spin properties, with valley pseudospin reproducing many optical features of spin materials. For example, valley-polarized excitons and carriers can be optically initialized using polarized light, mimicking the selection rules required for OISO~\cite{Zeng2012, Mak2012, Ye2017}. Despite this analogy, TMDs are not as optimal for electronic or spin device applications as many traditional semiconductors. Mobilities in TMDs are orders of magnitude smaller~\cite{Fuhrer2013} than traditional semiconductor spin-based devices~\cite{Mnatsakanov2004}, and spin-valley locking can impede free pseudospin manipulation by requiring unwieldy magnetic fields~\cite{Mitioglu2015,Plechinger2016}.

Other 2D semiconductors can present more favorable electronic and spin properties while preserving the benefits of layered materials.  Group-III monochalcogenides, such as GaSe and InSe, have dersiable electronic and magnetic properties that can persist in very thin layers~\cite{Li2015,Do2015,Premasiri2019,Song2020}. This broader class of materials offers a fresh platform for optical spin-based devices that combines layer-by-layer engineering with the ability to orient and freely manipulate a spin using polarized light that has long been exploited in traditional semiconductors.

Due to several noteworthy optoelectronic properties, InSe has gained significant attention over the last few years. In a few-layer InSe device at low temperature, the electron mobility can reach above $10^{4}$ cm$^{2}$/(Vs) and the quantum Hall effect can be observed~\cite{Bandurin2017}. Like TMDs, InSe has a layer-dependent band gap and relatively tightly bound ($\sim$ 10 meV) optically excitable and emissive excitonic states near the band edge~\cite{Bandurin2017,Song2020,Venanzi2020,Shubina2019}.  Unlike TMDs, the InSe direct band gap is near the $\Gamma$ point~\cite{Bandurin2017,Magorrian2016, Magorrian2017}, avoiding spin-valley locking caused by spin and off-center momentum correlations~\cite{Dey2017}. InSe is also predicted to have spin-dependent optical selection rules at these transition points for both the monolayer and multilayers~\cite{Magorrian2016, Magorrian2017}. This combination of properties suggests that InSe possesses the potential for 2D engineering of layered heterostructures while offering a distinct electronic and spin landscape that differs from TMDs.  Even though seminal demonstrations of optical spin phenomena have been reported in GaSe~\cite{Tang2015,Gamarts1977a}, direct experimental evidence of these optical spin phenomena in InSe, which has more favorable electronic properties than GaSe such as higher electron mobility and better on/off ratios for photodetection~\cite{Arora2021,Late2012}, is currently lacking.

Here, we present layer-sensitive OISO and spin dynamics in few and many-layered InSe. Polarized photoluminescence (PL) reveals both OISO and a layer-dependent emission polarization. Optical orientation and Zeeman splitting contributions to polarization can be identified separately in polarized PL. Although polarized PL is not observable in thicker InSe, time-resolved Kerr rotation (TRKR) reveals OISO persists in many-layer InSe. Spin precession in a magnetic field of optically oriented spin polarization is observed in thick InSe. Both polarized PL and TRKR reveal an effective magnetic moment less than expected for a free electron spin, yet consistent with the optically relevant spin phenomena originating from the strongly bound excitons in InSe~\cite{Shubina2019,Venanzi2020}. These results provide new insights to the spin properties of InSe and further establish the potential for layer-dependent orientation and manipulation of spin in InSe, opening the door for combining layer-sensitive spin properties with enticing electronic properties of 2D materials.

\section{Optically-Induced Spin Orientation in InSe}

In semiconductors, optical spin selection rules couple polarization of absorbed light with carrier spin polarization. For group-III monochalcogenide layered semiconductors, GaSe has been the primary case study for these optical selection rules in both theory~\cite{Li2015,Ivchenko1977} and experiment~\cite{Tang2015,Gamarts1977a}. The related layered semiconductor, InSe, has a similar band structure to GaSe and has also been predicted to have spin selection rules~\cite{Magorrian2016, Magorrian2017}. Because it has more favorable electronic properties than GaSe, such as a field effect mobility that is four orders of magnitude larger~\cite{Arora2021}, understanding the optical spin properties of carriers in InSe takes on practical significance for potential spintronics applications.
\begin{figure}[th]
\centering
\includegraphics[scale=1]{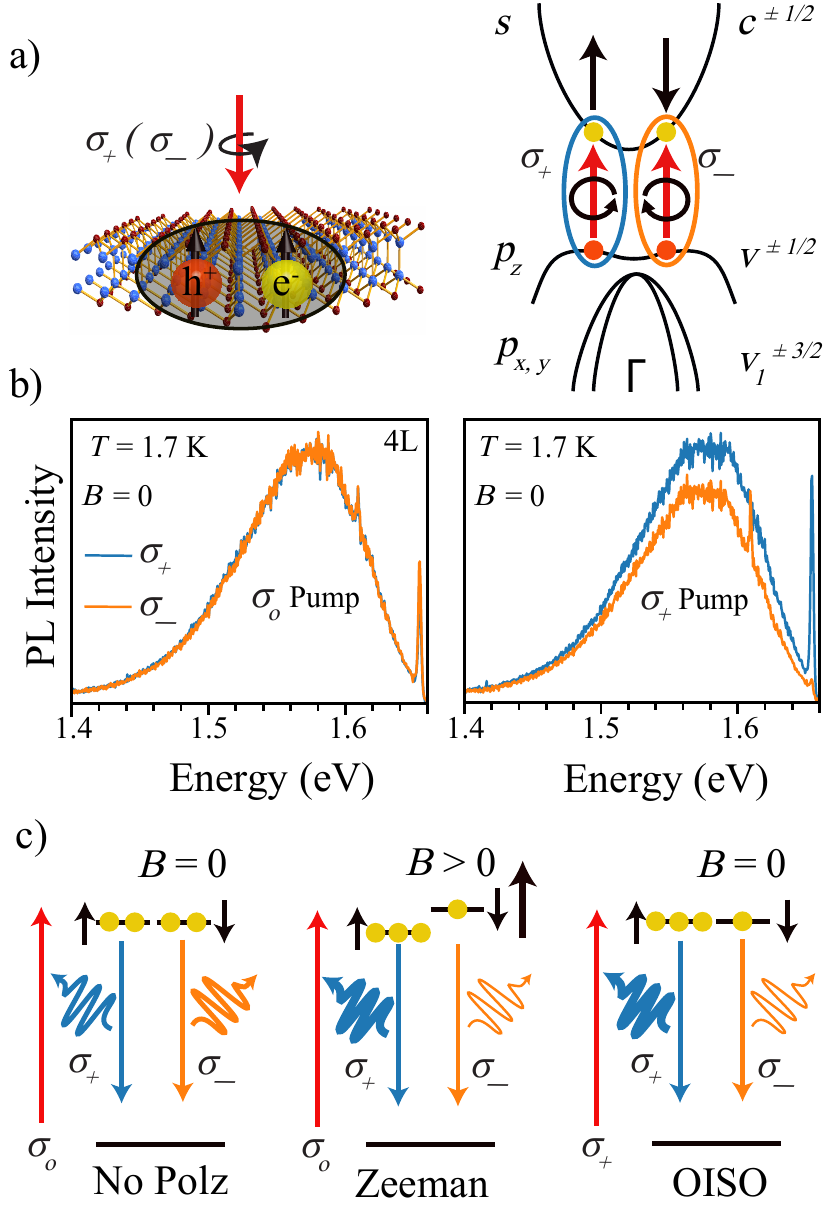}
\caption{ a) Illustration of a monolayer InSe crystal (left) and monolayer InSe band diagram (right). Monolayer and few-layer InSe has been predicted to have polarization dependent optical selection rules for band carriers and optical coupling to bound excitons due to strong binding energies.
b) Detection of polarization-resolved PL for 4L InSe for a linear (left) and circularly-polarized (right) pump.  The sharp peaks are from the tail of the pump laser centered at 1.67 eV.  c) Illustrations of several mechanisms underlying polarization-sensitive PL detection of spin-sensitive optical relaxation for different pump polarization and energy level configurations.  }
\label{fig:PPL}
\end{figure}

\begin{figure*}[tb]
\centering
\includegraphics[scale=1]{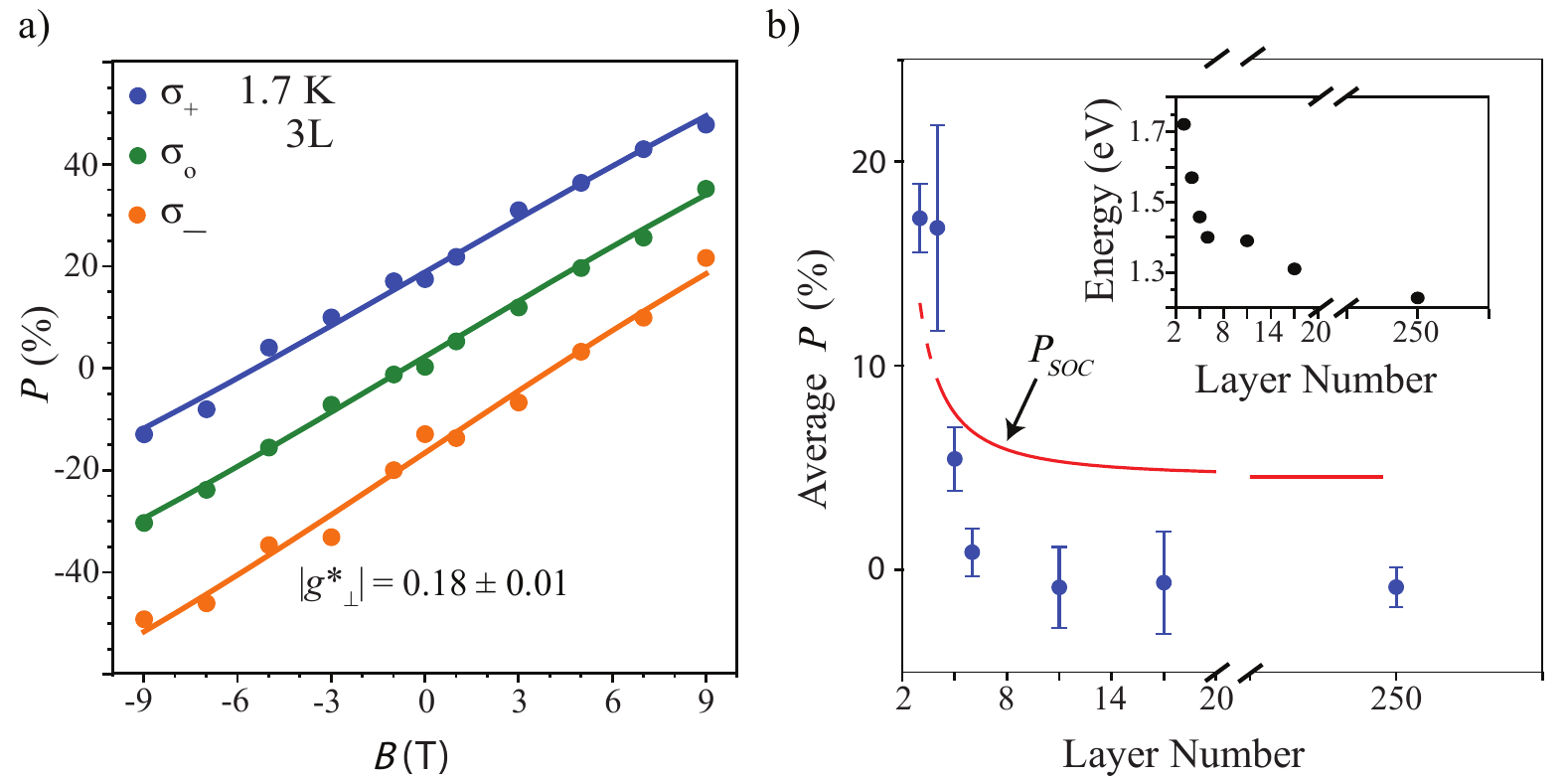}
\caption{ In a), $P$ vs B (magnetic field) for 3L InSe is shown. The linear trend for all excitations occurs due to Zeeman splitting whereas the offset of $\sigma_{+(-)}$ from $\sigma_{o}$, is due to OISO. Fitting these trends using Boltzmann statistics, we extract an effective $g$-factor. The average $P$ vs Layer Number is plotted in b). A qualitative fit is used to account for the Dresselhaus effect which gives a dependence of spin lifetime with layer number. The Dresselhaus term strictly holds for N $\geq$ 4L~\cite{Ceferino2021}. The inset directly relates peak of excitonic PL emission energy to thickness.}
\label{fig:OISO}
\end{figure*}

In InSe, the spin of excited carriers are coupled to circularly-polarized light absorbed at energies near the band-edge (illustrated in Fig.~\ref{fig:PPL}a). This interaction between spin and light  in InSe arises from the symmetry characteristics of the electronic band structure and their mixing~\cite{Magorrian2017}.
In the simplest model, near the $\Gamma$ point of few-layer InSe, the band-edge carrier wavefunctions are nominally a $s$-symmetry conduction band and $p_z$ orbital top valence band, suggesting no significant coupling to spin. Orbitals with $p_x$ and $p_y$ symmetry, which would couple with circularly-polarized optical fields, contribute to lower-energy valence bands. However, atomic spin-orbit coupling (SOC) disturbs this simple model by causing hybridization of the valence bands, thereby allowing spin-polarized optical transitions, by free carriers, from circularly-polarized light near the band gap energy~\cite{Magorrian2017}. This simple picture of free carrier optical transitions is complicated by the strongly bound excitons in InSe. Typical excitonic binding energies of $\sim$ 10 meV~\cite{Shubina2019, Venanzi2020} mean that at both room and cryogenic temperatures, optical phenomena are dictated by coupled dynamics of both electron and hole. Although the free carrier model highlights the main optical features of InSe, the excitonic effects should not be ignored when studying optical spin dynamics.

The transition selection rules contributing to band-edge excitons~\cite{KURODA1980,Brotons-Gisbert2019} suggests that spin polarization of InSe~\cite{Magorrian2017} can be observed from exciton photoluminescence (PL) emitted when excited by circularly polarized light. Opposite circular polarizations couple to band-edge excitons composed of the appropriately polarized electron and hole spin states, characterized by the exciton spin vector (Fig.~\ref{fig:PPL}a). When excited by circularly-polarized light, the degree of circular polarization ($P$) of the emitted exciton PL is a readout of the exciton spin population during the emission process. This expected optically-induced spin polarization is observed in band-edge excitonic PL at low temperature from 4-layer (4L) InSe prepared by mechanical exfoliation (Fig.~\ref{fig:PPL}b). The circular polarization-resolved PL shows that a $\sigma_{+}$ ($\sigma_{-}$) excitation beam generates $\sigma_{+}$ ($\sigma_{-}$) PL polarization even in the absence of a magnetic field, whereas a $\sigma_{o}$ (linear polarization) excitation results in unpolarized emission.

To illustrate  the possible mechanisms of the observed PL polarization, three PL detection scenarios are compared in Fig.~\ref{fig:PPL}c.  With linearly polarized excitation and no applied magnetic field, the spin-labeled excitonic levels are degenerate with no population imbalance, so no net polarization is detected (No Polz). With an applied magnetic field, linearly polarized excitation leads to polarized emission due to thermal relaxation to exciton levels split by the Zeeman effect (Zeeman). In the absence of an applied magnetic field, a circularly polarized pump will lead to spin-polarized exciton population imbalance resulting in net emission polarization (OISO).
The Zeeman and OISO scenarios are distinct mechanisms both leading to  PL polarization. Because of the absence of an applied magnetic field, in the right image of Fig.~\ref{fig:PPL}b, the results are due to the OISO response, originating strictly from the polarization-dependent exciton selection rules in InSe.

To explore the layer dependence of OISO in InSe, polarized PL measurements were conducted on exfoliated n-type InSe flakes of different thicknesses. Few-layer (3L - 6L) and multilayer ($<$ 20L) were encapsulated in hexagonal boron nitride (hBN) using standard layer transfer processes~\cite{Wang2013,Kretinin2014} in order to protect these samples from degrading while in ambient conditions. The thickest sample ($>$ 200L) was not encapsulated because bulk InSe does not degrade rapidly in air. To account for the layer-dependent band gap~\cite{Bandurin2017,Magorrian2016,Magorrian2017}, a tunable CW laser was used to pump near resonance for each InSe thickness. Excitation energies were less than 200 meV from resonance. Since the valence band spacing in few-layer InSe is on the order of $\sim$ 1 eV for layers greater than 2L~\cite{Bandurin2017,Magorrian2016}, in these measurements, the tunable laser allows excitation of just the lowest energy exciton level. PL energies for different InSe thickness is plotted in the inset of Fig.~\ref{fig:OISO}b. Polarized PL measurements were conducted in a magneto-optical helium-exchange cryostat. Further experimental setup details can be found in the Supplementary Information.

 The polarization, $P$, of the PL is calculated as percent polarization of the full spectrum, $(I_{+} - I_{-})/(I_{+} + I_{-})$. $I_{+(-)}$ is the intensity integrated over the polarized emission spectrum of interest. For different incident polarization on few-layer InSe, $P$ is measured at different magnetic fields (Fig.~\ref{fig:OISO}a). 
 All three excitation polarizations show a linear trend in $P$ with magnetic field $B$. The slope of each of these polarization trends are similar. These trends can be explained as arising from the relaxation and thermal equilibration in the Zeeman-split spin levels (Fig.~\ref{fig:PPL}c). The linear trends for the two circular polarizations have offsets of the same magnitude ($\sim$16\%) and opposite sign, demonstrating the expected effect of OISO. The $B$ response for $\sigma_{o}$ excitation is caused only by the Zeeman effect, while the  $\sigma_{+}$ and $\sigma_{-}$ response is the result of both OISO and the Zeeman effect.

 Since the linear trend is independent from OISO and is measured by the $\sigma_{o}$ excitation, it can be subtracted from the polarized excitation data. The remaining $P$ for both $\sigma_{+}$ and $\sigma_{-}$ are averaged and compared for samples with different layer numbers to isolate the dependence on thickness (Fig.~\ref{fig:OISO}b). The polarization from OISO is large for thin InSe ($<$ 5L) but decreases precipitously with more layers, indicating that optical spin polarization is a highly layer dependent phenomenon in InSe.

The origin of the layer-dependent polarization can be understood from the spin and recombination dynamics in InSe. Emission polarization is highly dependent on recombination and spin dynamics, parameterized in a simple rate model as:
\begin{equation}\label{eq:pol}
P=\frac{P_{0}}{1+\tau_{r}/\tau_{s}},
\end{equation}
where $P_{0}$ is the degree of circular polarization at the time of excitation and $\tau_r$ and $\tau_s$ are the recombination and spin lifetimes, respectively~\cite{Zutic2004}. Calculations show that absorption of in-plane circularly polarized light does not change significantly with thickness in InSe~\cite{Magorrian2016,Magorrian2017}. Thus, assuming $P_{0}$ is constant with thickness, our observations suggest that the ratio $\tau_{r}/\tau_{s}$ increases with layer number. This dependence impacts the ability to observe OISO in multilayers using polarized PL and it is a central component of the discussion of optical spin phenomena later in this manuscript.

\begin{figure*}[tb]
\centering
\includegraphics[scale=1]{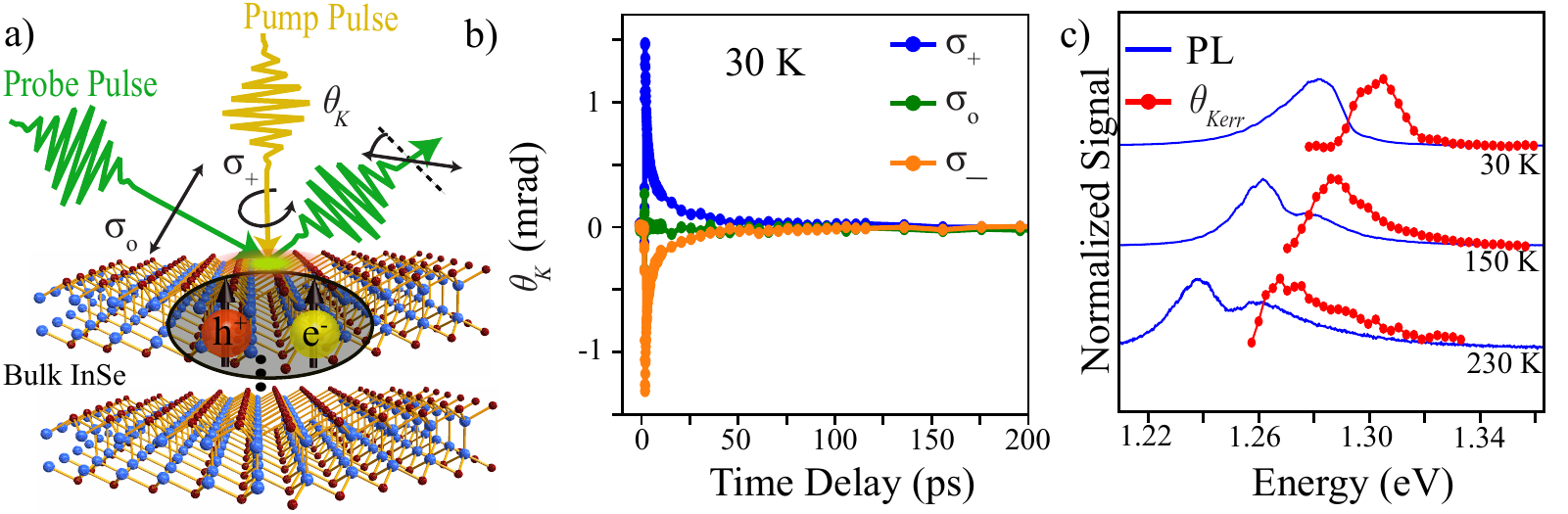}
\caption{ A cartoon of the TRKR method and TRKR data are shown for bulk InSe in a) and b) respectively. Kerr signal ($\theta_{\rm K}$) was maximized for a probe photon energy of 1.30 eV. The spin lifetime in bulk InSe was extracted to be $\sim$ 25 ps. A comparison of wavelength dependent $\theta_{\rm K}$ (at 3 ps) to PL for different temperatures is shown in c).}
\label{fig:TRKR}
\end{figure*}

The magnitude of the effective out-of-plane $g$-factor, ($\abs{g^{*}_{\perp}}$) can be estimated from the linear trend in $P$ caused by the Zeeman splitting. Assuming that excited carriers thermally relax before emission, a magnetic field-induced spin splitting will lead to a thermal spin imbalance that manifests in net polarization (Fig.~\ref{fig:PPL}c). The net emission polarization is a measure of the net spin polarization, which comes from Boltzmann statistics as:
\begin{equation}\label{eq:zeeman}
P = \tanh{\left[{\frac{\mu_{B}\abs{g^{*}_{\perp}}B}{2k_{b}T}}\right]},
\end{equation}
where $\mu_{B}$ is the Bohr magneton and $T$ is temperature. At low magnetic fields, $P$ is linear with magnetic field, with slope related to the $g$-factor. For 3L InSe, $\abs{g^{*}_{\perp}}\approx 0.2$. The Supplementary Information contains extracted values for other thicknesses. All values sit between 0.08 and 0.24. For the parameters of the experiment, this model (Eq.~\ref{eq:zeeman}) is in the linear regime of magnetic field. Stronger evidence of this Zeeman model  would be provided by higher magnetic fields not accessible with the existing instrumentation.

The $g$-factor of monochalcogenides has been treated in prior studies. Li and Applebaum theoretically predict that $\abs{g^{*}_{\perp}} <$ 2 for nearly-free electrons in GaSe~\cite{Li2015}. Bandurin \textit{et al.} estimate a $g$-factor for n-type InSe using transport measurements, putting the effective $g$-factor closer to 2, the expected value for a free electron~\cite{Bandurin2017}. These results are not directly comparable to $g$-factors extracted from OISO because optical processes in InSe are dominated by excitons with large binding energies ($\sim$ 10 meV)~\cite{Shubina2019, Venanzi2020} and long emissive lifetimes ($\sim$ 1 ns)~\cite{Venanzi2020}. Many experiments explicitly consider excitons when studying optical properties of InSe ~\cite{Bandurin2017,Shubina2019,Venanzi2020,Zhong2019,Kuroda1982}, however, theoretical literature that explicitly deals with roles excitonic states play in optical spin properties in InSe is lacking.

\section{Spin Dynamics in Many-Layer InSe}
 Because of the rapid decrease in PL polarization with layer number, thick InSe, including bulk crystals, do not show evidence of OISO in emission. Yet, this does not rule out the existence of OISO in thick InSe. If the lifetime ratio in Eq.~\ref{eq:pol} increases with layer number, as would be expected if spin lifetimes decrease, then emission polarization is suppressed even with robust OISO during excitation. This implies that polarized PL is not a sensitive probe for detecting OISO for thicker InSe.  Time-resolved studies show that spin polarization in GaSe persists in nanoslabs with thicknesses greater than 100 nm~\cite{Tang2015}. Therefore, we use time-resolved Kerr rotation (TRKR) to detect spin polarization in many-layer InSe.

TRKR is an ultrafast pump-probe technique widely used to study spin properties in III-V semiconductors~\cite{Press2008,Yugova2007,Yang2010,Rudolph2015} and TMDs~\cite{Dey2017,Yang2015,McCormick2017,Zhu2014, LaMountain2018} because of its sensitivity to spin polarization. A circularly polarized pulse optically pumps a non-equilibrium spin imbalance. As the spin population relaxes to equilibrium, a linearly polarized probe pulse, with a time delay relative to the pump, monitors the relaxation through the Kerr rotation angle $\theta_{\rm K}$ of its linear polarization axis.

We used a one-color pump-probe scheme in the near infrared (1.26 eV - 1.36 eV) with a ratio of pump to probe power of 10:1. The InSe sample was placed in a closed-cycle magneto-optical cryostat designed for free space optics. Due to spot size limitations, experiments are done on InSe with thickness $\geq$ 500 nm to achieve a sample flake size of about 60 $\mu$m $\times$ 60 $\mu$m. The Supplementary Information contains additional TRKR experimental details.

TRKR is observed in thick InSe with $\theta_{\rm K}$ vs Time Delay for the three pump helicities (Fig.~\ref{fig:TRKR}). Kerr signals with similar magnitude and opposite rotation are observed for the $\sigma_{+}$ and $\sigma_{-}$ pump, indicating opposite polarity of spin population as expected. For a linear pump, a symmetric spin population is produced and there is no signal. Kerr rotation spectral features of InSe coincide roughly with excitonic absorption~\cite{Venanzi2020,Shubina2019}, although the peak TRKR signal differs from the absorption or emission energy.  Comparing the TRKR spectrum at time delay $\sim$ 3 ps to the PL emission spectra reveals a good correlation between energies across temperatures (Fig.~\ref{fig:TRKR}c), supporting that TRKR originates from band-edge exctons.

\begin{figure}[ht]
\centering
\includegraphics[scale=1]{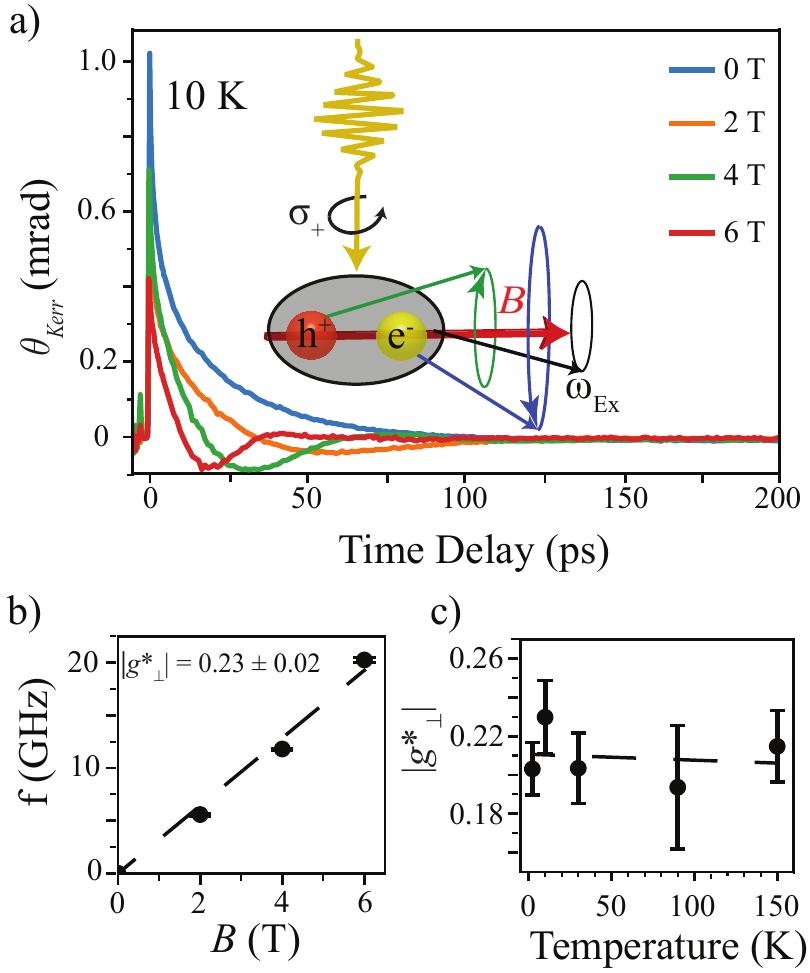}
\caption{a) Kerr rotation for thick InSe in a perpendicular magnetic field, revealing oscillations of the excitonic spin orientation. Inset shows a schematic of the circular pump excitation causing the precession of exciton spin in the presence on in-plane magnetic field b) The Larmor spin frequency at $T = 10$~K is linear with magnetic field $B$. c) The effective $g$-factor does not change significantly for temperatures between 4 K and 150 K.  }
\label{fig:Prec}
\end{figure}

Applying a transverse magnetic field to the optically-excited spin population leads to spin precession. As the field increases, oscillations within short lifetimes ($\sim$ 25 ps) appear in the TRKR signal (Fig.~\ref{fig:Prec}a), evident most clearly at high fields (high oscillation frequencies). At the experimentally accessible fields, only a few oscillations can be observed in the relaxation lifetime. Therefore, the spin precession is analyzed using a phenomenological model of an exponentially decaying cosine function, $\theta_{\rm{K}} \propto e^{-t/\tau_{s}}\cos{[(2\pi f)t]}$, where $f$ is the Larmor frequency and $\tau_{s}$ is the spin lifetime. $\abs{g^{*}_{\perp}}$ is extracted from the frequency field dependence, $f = \frac{\mu_{B}gB}{4 \pi}$, where $\mu_{B}$ is the Bohr magneton and $B$ is the applied  magnetic field.

A fit to $f$ vs $B$ yields an effective $g$-factor, $\abs{g^{*}_{\perp}}$ = 0.23 $\pm$ 0.02 (Fig.~\ref{fig:Prec}b). This $g$-factor is similar to that obtained from the Zeeman model for polarized PL in thin InSe. TRKR measurements are performed on bulk InSe across a wide temperature range (Fig.~\ref{fig:Prec}c). At each temperature, the photon energy was set to maximize the Kerr signal. Below $T = 150$ K, $\abs{g^{*}_{\perp}}$ is essentially unchanged. Higher temperatures are not explored because of reduced signal-to-noise.

\section{Discussion}

These results reveal direct evidence of OISO in both few-layer and bulk InSe. Although the conclusion of optical spin orientation is well supported, the measurements reveal two key observations about InSe OISO dynamics in need of explanation: (1) the persistence of OISO in bulk for short timescales in TRKR, even though steady state polarized PL shows \textit{P} approaching zero for thicknesses greater  than 5L, and (2) the $g$-factor of $\sim$ 0.2, which is  relatively consistent across thickness regimes, yet
much smaller than expected value for highly mobile electrons in InSe~\cite{Bandurin2017}.

To address the first issue, we consider the expected impact of layer number on OISO in InSe. Although there are predictions of OISO in monochalcogenides such as GaSe~\cite{Li2015, Ivchenko1977} and InSe~\cite{Magorrian2016,Magorrian2017}, prior experimental observations of OISO are only in bulk GaSe~\cite{Tang2015,Gamarts1977a}. The results here in few-layer and many-layer InSe present a much broader exploration of this phenomenon. The band structure of InSe is highly layer-dependent. The primary band gap changes non-linearly with layer number, but the shape of the conduction and valence bands allow for optical transitions near the $\Gamma$ point for any thickness~\cite{Magorrian2016,Rybkovskiy2014}. For monolayer InSe, the valence band is shaped like a caldera, with maxima not far from the $\Gamma$ point. For thicker layers, the valence band is relatively flat. The conduction band, on the other hand, remains similar in shape for different thicknesses~\cite{ Magorrian2016,Rybkovskiy2014}. The absorption strength for in-plane circularly polarized light is not expected to depend significantly on thickness~\cite{Magorrian2017}, suggesting similar optical spin pumping efficiency independent of thickness. These predictions suggest that band shape and strength of polarization selection rules for optical transitions play little role in determining layer-dependent polarization of emission.  Therefore, we look towards relaxation mechanisms as the key to understanding the layer-dependence.

Prior work has considered spin relaxation mechanisms in group-III monochalcogenides~\cite{Li2015,Tang2015,Do2015,Ceferino2021,Takasuna2017}. SOC can exist in these materials for noncentrosymmetric polytypes~\cite{Li2015,Do2015} (Dresselhaus effect) or when breaking mirror-plane symmetry with an induced electric field (Rashba effect) for thicknesses greater than a monolayer~\cite{Li2015,Ceferino2021}. 
Both types of symmetry-dependent SOC affect the D'yakonov-Perel (DP) spin relaxation~\cite{Li2015,Do2015,Ceferino2021}. Although it is not strongly dependent on SOC, the Elliot-Yafet (EY) spin relaxation can also play a major role in monochalcogenide spin dynamics~\cite{Li2015, Tang2015}. Experiments on GaSe suggest that at low temperatures, DP relaxation dominates in thin samples~\cite{Takasuna2017}, while EY relaxation has been argued to dominate in thicker samples~\cite{Tang2015}. Overall, these results imply that spin relaxation in monochalcogenides is sensitive to thickness. In contrast, experiments have shown that the recombination lifetime of election-hole pairs is on the order of $\sim$1 ns and relatively insensitive to layer number~\cite{Venanzi2020,Tang2015}.

These trends can be roughly confirmed in our InSe samples. For thicknesses $\geq$ 500 nm, ultrafast pump-probe experiments provides an estimate of $\tau_{r}\sim$ 1 ns from the time-resolved reflectance using cross-polarized linear beams (see Supplementary Information). TRKR suggests $\tau_{s}\sim$ 25 ps. Thus,  $\tau_{s}\ll\tau_{r}$ would make OISO unobservable for thick samples in polarized PL measurements. Unfortunately, due to constraints with the time-resolved apparatus, we cannot measure both $\tau_{r}$ and $\tau_{s}$ independently for thin layers for comparison.

Assuming the DP mechanism dominates relaxation for thin samples and knowing that the momentum correlation time for an electron ($\tau_{p}$) is less than 1 ps  based on mobility measurements~\cite{Bandurin2017}, the spin relaxation is in the motional narrowing regime~\cite{Wu2010}. This implies that 1/$\tau_{s}$ = $\Omega^{2}\tau_{p}$, where $\Omega$ is the effective SOC precession frequency, a property of the band structure. This can be written as $\Omega = \alpha k_{F}$ where $\alpha$ is the strength of the effective field originating from SOC and $k_{F}$ is the Fermi momentum. $\alpha$ is layer dependent and depends on the underlying SOC mechanisms, with $\alpha (N) \propto \alpha_{D}(N)$~\cite{Ceferino2021}. Here, $\alpha_{D}$ is the SOC strength for Dresselhaus effect and $N$ is the number of layers. The functional form of $\alpha_{D}$ can be written as~\cite{Ceferino2021}:
\begin{equation}\label{SOC}
\centering
 \alpha_{D}(N)\approx\alpha_{\infty}\left (  1 -\frac{\chi}{(N+2.84)^2} \right )
\end{equation}
where $\alpha_{\infty}$ and $\chi$ are constants that account for bulk properties. The first describes the SOC at the conduction band edge and the second represents the SOC having nonlinear dependence. The relationship of this parameter with spin lifetimes is $1/\tau_{s} (N) \approx A\alpha_{D}^{2}(N)$, where $A$ is a constant. Inserting this parameterization into Eq.~\ref{eq:pol} gives a simple model relating emission polarization $P$ to layer number $N$. Since $\alpha_{\infty}$ and $\chi$ are previously determined for InSe~\cite{Ceferino2021}, and $\tau_{r}$ is estimated from time-resolved data, the only undetermined parameters in this model are the proportionality constants $P_{0}$ and $A$. A more detailed discussion of the SOC model from Ref.~\onlinecite{Ceferino2021} and the layer-dependent model for $P$ is in the Supplementary Information.

The layer-dependent OISO data is fit to this model using $P_{0}$ and A as the only free fit parameters (Fig.~\ref{fig:OISO}b). Although this simple formulation is not quantitatively accurate, the model captures the swiftly decreasing trend of $P$ with layer number. The main physics encoded in this model is that 1/$\tau_{s}$ increases with thickness, which matches the trends in the experimental data and expectations for $\tau_{s}$ from literature~\cite{Ceferino2021}. This model, however, neglects spin-spin scattering. Although predictions expect this effect to be negligible~\cite{Do2015}, experimental work suggests otherwise~\cite{Tang2015}. Thus, the EY mechanism likely needs to be included in a full model of layer dependent spin relaxation in group-III monochalcogenides. Even so, the current simple model suggested here captures the experimental trends and provides some insight into the detailed mechanisms of the OISO observed in polarized PL.

The second key observation from the OISO and spin dynamics data is the value of $g\sim 0.2$. Although both polarized PL and TRKR are required to explore the full range of InSe thicknesses, both methods reveal similar $g$-factors. This measurement is distinct from the value expected for free electrons in InSe. As explained previously, prior observations estimate $g \sim 2$  for group III monochalcogenides~\cite{Li2015,Bandurin2017}. Currently in the literature, measurements and predictions of $g$-factor have looked solely at electron carriers. Because of the strong excitonic binding, the full explanation for the observed $g$-factor of optically excited spin polarization and its insensitivity to layer number must incorporate excitonic effects.

\begin{figure}[htb]
\centering
\includegraphics[scale=1]{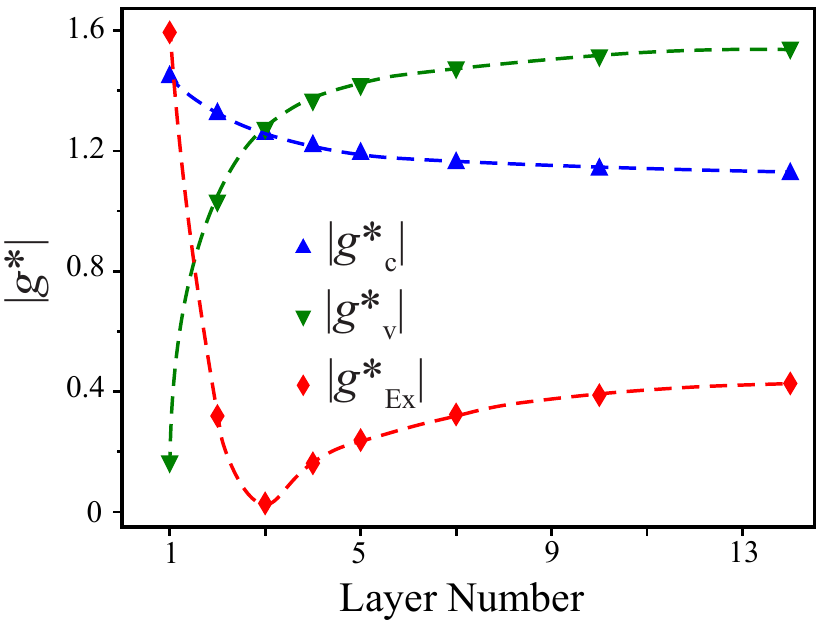}
\caption{ Calculated magnitude of out-of-plane $g$-factors for electrons $(g^{*}_{\rm c}$), holes ($g^{*}_{\rm v}$), and excitons ($g^{*}_{\rm Ex}$) using a few-layer band model adapted from Ref.~\onlinecite{Magorrian2017}. The dashed lines are guides to the eye. }
\label{fig:g-factor}
\end{figure}

 A recent fully-parameterized theoretical framework developed to model InSe optical transitions~\cite{Magorrian2017} can be adapted to model the $g$-factor in InSe multilayers.  As explained in Ref.~\onlinecite{Magorrian2017}, a $\boldsymbol{k} \cdot \boldsymbol{p}$ perturbation theory Hamiltonian for monolayer InSe can be constructed from the basis of atomic orbitals, with off-diagonal couplings originating from momentum matrix elements and SOC. This layer Hamiltonian can then be ``stacked’’ using a hopping model to give a multilayer Hamiltonian.  From the zone-center band solutions of this Hamiltonian and appropriate momentum matrix elements, the orbital $g$-factor can be calculated separately for conduction and valence bands~\cite{Hermann1977,Wang2015}. Using these values, the effective out-of-plane $g$-factor is calculated using $g^{*}_{ \rm c,v} = 2 + g^{\rm orb}_{ \rm c,v}$. Where the first term is the $g$-factor for a free electron (hole) and the second term represents the orbital $g$-factor for the conduction band or valence band. Li and Applebaum calculate $\abs{g^{\rm orb}_{c}} =  0.3$ for monolayer GaSe~\cite{Li2015}. Our band model gives $\abs{g^{orb}_{c}}\approx 0.5$ for monolayer InSe, which is of a similar magnitude. Thickness dependence for the magnitude of $g^{*}_{ \rm c,v}$ can be seen in Fig.~\ref{fig:g-factor}. Only the magnitude is considered for calculated $g$-factors since our experiments cannot determine the sign.

This framework provides a  reasonable basis to model excitonic $g$-factors in InSe. The exciton $g$-factor ($g^{*}_{\rm Ex}$) can be obtained as $ g^{*}_{\rm Ex} =  g^{*}_{\rm c}$ – $g^{*}_{\rm v}$. For layers between 3L to 14L, the magnitude of the effective exciton $g$-factor sits between 0 and 0.5 (Fig.~\ref{fig:g-factor}). These values obtained from this basic $\boldsymbol{k}\cdot \boldsymbol{p}$ multilayer parameterization are close to the those from experiment with very little layer dependence, and they are distinct from the free carrier expectation~\cite{Li2015,Bandurin2017}.

\begin{figure}[ht]
\centering
\includegraphics[scale=1]{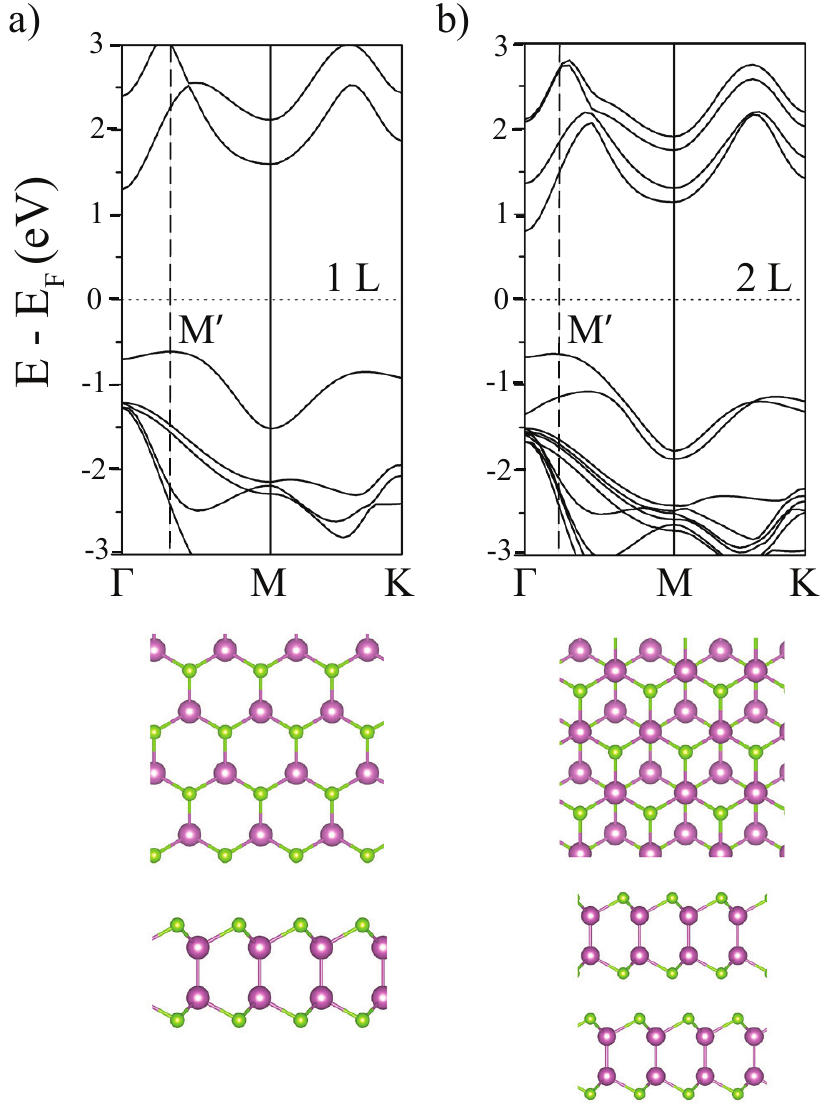}
\caption{PBE band structures for a) a monolayer and b) AB stacked bilayer InSe. Top and side view of a) monolayer and b) AB stacking bilayer InSe can seen at the bottom of the figure. Purple and green spheres represent In and Se atoms, respectively. }
\label{fig:DFT}
\end{figure}

The $\boldsymbol{k}\cdot \boldsymbol{p}$ model is best thought of as a toy model to formulate basic expectations rather than a rigorously accurate calculation since excitonic effects, layer effects, and details of real materials are not accounted for in extensive depth.  To provide additional and independent support for the results of this toy model, we have computed the exciton $g$-factor for  monolayer and bilayer InSe using {\it ab-initio} Density Functional Theory (DFT) methods.
For a bilayer structure, we considered AB stacking (Fig.~\ref{fig:DFT}b, bottom) as a previous DFT study showed that the AB stacking mode is more favorable than the AA stacking mode~\cite{yao21}.
The $g$-factor for band $n$ at wave vector $k$ can be estimated by~\cite{forste20,forg21}
\begin{equation}
g_{n}(\boldsymbol{k})=g_{0}s_{n}+2L_{n}(\boldsymbol{k})
\end{equation}
where $g_{0}$, $s_{n}$, and $L_{n}(\boldsymbol{k})$ represent the free electron Land\'e factor $g_{0} \sim 2$, spin projection, and orbital angular momentum, respectively. The $z$-component of the orbital angular momentum can be written as~\cite{roth59,xiao10}

\begin{equation}
L_{n}(\boldsymbol{k}) = \frac{2m_{0}}{\hbar^{2}}\sum_{m \neq n}\text{Im}[\xi_{nm}^{(x)}(\boldsymbol{k})\xi_{mn}^{(y)}(\boldsymbol{k})](E_{n\boldsymbol{k}}-E_{m\boldsymbol{k}}).
\end{equation}
Here, $E_{n\boldsymbol{k}}$ and $\boldsymbol{\xi}_{nm}(\boldsymbol{k})= i \langle u_{n\boldsymbol{k}}|\partial/\partial\boldsymbol{k}|u_{m\boldsymbol{k}} \rangle$ indicate the eigenvalue of band $n$ at wave vector $\boldsymbol{k}$ and the interband matrix of the coordinate operator, respectively;
$u_{n\boldsymbol{k}}$ is the periodic part of the Bloch wavefunction which can be obtained numerically in DFT scheme.
The DFT calculations were performed within the density functional approximation using the Generalized Gradient Approximation exchange-correlation functional with the Perdew-Burke-Ernzerhof (PBE) parametrization~\cite{perdew97} as implemented in the Quantum Espresso package~\cite{giannozzi09}.
The DFT calculations were done with a plane-wave basis set with a 400~Ry kinetic energy cut-off for the energy-consistent norm-conserving Burkatzki, Filippi, and Dolg (BFD) pseudopotentials~\cite{burkatzki07,burkatzki08} for In and Se, and $10\times10\times1$ $k$ grids used for both mono- and bilayer InSe.
In order to incorporate van der Waals interlayer interactions within the DFT framework, the dispersion of Grimme (DFT-D3)~\cite{grimme10} is applied for bilayer InSe.
Using DFT-D3, the equilibrium interlayer distance of AB-stacked bilayer InSe is estimated to be  3.2~\AA.

Figure~\ref{fig:DFT} shows the calculated band structure of layered InSe.
We see that valence band maximum (VBM) is located between the $\Gamma$ and M high symmetry points, but only $\sim$ 0.1 eV energy difference is shown between indirect and direct gaps at the $\Gamma$ point both for monolayer and bilayer structures.
Computed band gaps and exciton g-factors are summarized in Tab.~\ref{tab:gfactor}.
In this study, we consider three high symmetry points ($\Gamma$, K, and M), and denote by M' the location of the VBM.

For $\Gamma \rightarrow \Gamma$, the $|g^{*}_{Ex}|$ for monolayer and bilayer are 1.3 and 0.1, respectively. Comparing these to values calculated from the toy model, 1.6 and 0.4, it can be seen that these results are on the same order of magnitude for respective layers. Even more important, as layer number goes from 1L to 2L, there is a significant decrease in the effective excitonic $g$-factor that is reflected in both theoretical methods. The similar trend with $g$-factors directly extracted from DFT gives credence to our first method and offers further confirmation of our observations of excitonic $g$-factors in InSe. Due to limitations of experimental apparatus (wavelengths, spot size, etc.), we can not experimentally investigate monolayer and bilayer, therefore direct experimental comparison to DFT calculations are left for future studies.

\begin{table}[t]
\centering
\caption{PBE band gap and exciton g-factor for a monolayer and bilayer InSe}
\label{tab:gfactor}
\begin{tabular}{c|cc|cccc}
\hline \hline
                        &  \multicolumn{2}{c|}{band gap (eV)}  &  \multicolumn{4}{c}{$|g^{*}_{Ex}|$} \\\
                        &  Direct      &   Indirect  &   $\Gamma \rightarrow \Gamma$     &  $M'$ $\rightarrow \Gamma$   &  $M$ $\rightarrow$ $M$    & $K$ $\rightarrow$ $K$  \\ \hline
 Monolayer      & 2.0     &  1.9      &   1.3   & 2.7  &   1.6  &  0.2  \\
 Bilayer &        1.5     &  1.4   &   0.1     &   0.9    & 0.0    &  0.3 \\ \hline \hline
\end{tabular}
\end{table}

Overall, this analysis confirms that the experiments are probing InSe exciton spin precession, consistent with expectations. Being based on a carrier band model, neither the toy $\boldsymbol{k}\cdot \boldsymbol{p}$ model nor the {\it ab-initio} DFT methods explicitly incorporate exciton effects in the band structure. It is likely that a more extensive theoretical framework is needed to better understand  the quantitative spin dynamics of excitons in InSe, as deeper valence bands are also predicted to have polarized selection rules~\cite{Magorrian2017}. The experimental results presented here suggest that such an excitonic spin dynamics theory is needed, but the present toy model and DFT calculations do provide convincing support for our interpretation of the data.

\section{Conclusion}

In this report, we have shown direct experimental observations of OISO and spin precession in InSe. The spin selection rules predicted for few-layer InSe are confirmed and shown to persist in bulk materials. The observed emission polarization from OISO is layer dependent and vanishes for thicknesses $>$ 5L, originating from layer-dependent spin relaxation. A Zeeman model and phenomenological spin precession model were used to extract $\abs{g^{*}_{\perp}} < $ 0.24 for various thickness of InSe. This spin precession rate differs from the free carrier rate and likely originates from the excitonic nature of optical transitions in InSe, which are neglected in common InSe band models of optical transitions.  These results mark the first foray into exploring optical spin dynamics and optical spin orientation in layered InSe, helping determine its viability as a layer-sensitive 2D platform for spin-based devices.

\begin{acknowledgments}
This material is based upon work primarily supported by the National Science Foundation under Grant No. DMR-1905986 and by the MRSEC program at Northwestern (DMR-1720319). J.N. was supported by the NSF Graduate Research Fellowship (DGE-1842165). D.L. thanks the Swiss National Science Foundation for an Early PostDoc Mobility Fellowship (P2EZP2\_181614) and the Materials Research Science and Engineering Center of Northwestern University (NSF DMR-1720139). This work made use of the Northwestern University NUANCE Center and the Northwestern University Micro/Nano Fabrication Facility (NUFAB), which have received support from the SHyNE Resource (NSF ECCS-1542205), the International Institute for Nanotechnology, and the Northwestern University MRSEC program (NSF DMR-1720139). The 2D crystal manipulation system (Graphene Industries) used in this work was supported by an Office of Naval Research DURIP grant (ONR N00014-19-1-2297). T.L. was supported by the Office of Naval Research under grant number N00014-16-1-3055. J.T.G. was supported by the National Science Foundation Division of Materials Research (NSF DMR-2004420).  H.S. and O.H. were supported by the U.S. Department of Energy, Office of Science, Basic Energy Sciences, Materials Sciences and Engineering Division, as part of the Computational Materials Sciences Program and Center for Predictive Simulation of Functional Materials. K.W. and T.T. acknowledge support from JSPS KAKENHI (Grant Numbers 19H05790, 20H00354 and 21H05233) for hBN synthesis.
An award of computer time was provided by the Innovative and Novel Computational Impact on Theory and Experiment (INCITE) program and was used to generate all DFT results. This research used resources of the Argonne Leadership Computing Facility, which is a DOE Office of Science User Facility supported under contract DE-AC02-06CH11357.

\end{acknowledgments}

\bibliography{Nelson2022}

\begin{thebibliography}{58}%
\makeatletter
\providecommand \@ifxundefined [1]{%
 \@ifx{#1\undefined}
}%
\providecommand \@ifnum [1]{%
 \ifnum #1\expandafter \@firstoftwo
 \else \expandafter \@secondoftwo
 \fi
}%
\providecommand \@ifx [1]{%
 \ifx #1\expandafter \@firstoftwo
 \else \expandafter \@secondoftwo
 \fi
}%
\providecommand \natexlab [1]{#1}%
\providecommand \enquote  [1]{``#1''}%
\providecommand \bibnamefont  [1]{#1}%
\providecommand \bibfnamefont [1]{#1}%
\providecommand \citenamefont [1]{#1}%
\providecommand \href@noop [0]{\@secondoftwo}%
\providecommand \href [0]{\begingroup \@sanitize@url \@href}%
\providecommand \@href[1]{\@@startlink{#1}\@@href}%
\providecommand \@@href[1]{\endgroup#1\@@endlink}%
\providecommand \@sanitize@url [0]{\catcode `\\12\catcode `\$12\catcode
  `\&12\catcode `\#12\catcode `\^12\catcode `\_12\catcode `\%12\relax}%
\providecommand \@@startlink[1]{}%
\providecommand \@@endlink[0]{}%
\providecommand \url  [0]{\begingroup\@sanitize@url \@url }%
\providecommand \@url [1]{\endgroup\@href {#1}{\urlprefix }}%
\providecommand \urlprefix  [0]{URL }%
\providecommand \Eprint [0]{\href }%
\providecommand \doibase [0]{https://doi.org/}%
\providecommand \selectlanguage [0]{\@gobble}%
\providecommand \bibinfo  [0]{\@secondoftwo}%
\providecommand \bibfield  [0]{\@secondoftwo}%
\providecommand \translation [1]{[#1]}%
\providecommand \BibitemOpen [0]{}%
\providecommand \bibitemStop [0]{}%
\providecommand \bibitemNoStop [0]{.\EOS\space}%
\providecommand \EOS [0]{\spacefactor3000\relax}%
\providecommand \BibitemShut  [1]{\csname bibitem#1\endcsname}%
\let\auto@bib@innerbib\@empty
\bibitem [{\citenamefont {\u{Z}uti\'{c}}\ \emph {et~al.}(2004)\citenamefont
  {\u{Z}uti\'{c}}, , \citenamefont {Fabian},\ and\ \citenamefont {{Das
  Sarma}}}]{Zutic2004}%
  \BibitemOpen
  \bibfield  {author} {\bibinfo {author} {\bibfnamefont {I.}~\bibnamefont
  {\u{Z}uti\'{c}}}, , \bibinfo {author} {\bibfnamefont {J.}~\bibnamefont
  {Fabian}},\ and\ \bibinfo {author} {\bibfnamefont {S.}~\bibnamefont {{Das
  Sarma}}},\ }\bibfield  {title} {\bibinfo {title} {{Spintronics:Fundamnetal
  and applications}},\ }\href@noop {} {\bibfield  {journal} {\bibinfo
  {journal} {Reviews of Modern Physics}\ }\textbf {\bibinfo {volume} {76}},\
  \bibinfo {pages} {323} (\bibinfo {year} {2004})}\BibitemShut {NoStop}%
\bibitem [{\citenamefont {Crooker}\ \emph {et~al.}(2007)\citenamefont
  {Crooker}, \citenamefont {Furis}, \citenamefont {Lou}, \citenamefont
  {Crowell}, \citenamefont {Smith}, \citenamefont {Adelmann},\ and\
  \citenamefont {Palmstr{\o}m}}]{Crooker2007}%
  \BibitemOpen
  \bibfield  {author} {\bibinfo {author} {\bibfnamefont {S.~A.}\ \bibnamefont
  {Crooker}}, \bibinfo {author} {\bibfnamefont {M.}~\bibnamefont {Furis}},
  \bibinfo {author} {\bibfnamefont {X.}~\bibnamefont {Lou}}, \bibinfo {author}
  {\bibfnamefont {P.~A.}\ \bibnamefont {Crowell}}, \bibinfo {author}
  {\bibfnamefont {D.~L.}\ \bibnamefont {Smith}}, \bibinfo {author}
  {\bibfnamefont {C.}~\bibnamefont {Adelmann}},\ and\ \bibinfo {author}
  {\bibfnamefont {C.~J.}\ \bibnamefont {Palmstr{\o}m}},\ }\bibfield  {title}
  {\bibinfo {title} {{Optical and electrical spin injection and spin transport
  in hybrid Fe/GaAs devices}},\ }\href {https://doi.org/10.1063/1.2722785}
  {\bibfield  {journal} {\bibinfo  {journal} {Journal of Applied Physics}\
  }\textbf {\bibinfo {volume} {101}},\ \bibinfo {pages} {1} (\bibinfo {year}
  {2007})}\BibitemShut {NoStop}%
\bibitem [{\citenamefont {Kroutvar}\ \emph {et~al.}(2004)\citenamefont
  {Kroutvar}, \citenamefont {Ducommun}, \citenamefont {Heiss}, \citenamefont
  {Bichler}, \citenamefont {Schuh}, \citenamefont {Abstreiter},\ and\
  \citenamefont {Finley}}]{Kroutvar2004}%
  \BibitemOpen
  \bibfield  {author} {\bibinfo {author} {\bibfnamefont {M.}~\bibnamefont
  {Kroutvar}}, \bibinfo {author} {\bibfnamefont {Y.}~\bibnamefont {Ducommun}},
  \bibinfo {author} {\bibfnamefont {D.}~\bibnamefont {Heiss}}, \bibinfo
  {author} {\bibfnamefont {M.}~\bibnamefont {Bichler}}, \bibinfo {author}
  {\bibfnamefont {D.}~\bibnamefont {Schuh}}, \bibinfo {author} {\bibfnamefont
  {G.}~\bibnamefont {Abstreiter}},\ and\ \bibinfo {author} {\bibfnamefont
  {J.~J.}\ \bibnamefont {Finley}},\ }\bibfield  {title} {\bibinfo {title}
  {{Optically programmable electron spin memory using semiconductor quantum
  dots}},\ }\href {https://doi.org/10.1038/nature03008} {\bibfield  {journal}
  {\bibinfo  {journal} {Nature}\ }\textbf {\bibinfo {volume} {432}},\ \bibinfo
  {pages} {81} (\bibinfo {year} {2004})}\BibitemShut {NoStop}%
\bibitem [{\citenamefont {Greilich}\ \emph {et~al.}(2006)\citenamefont
  {Greilich}, \citenamefont {Yakovlev}, \citenamefont {Shabaev}, \citenamefont
  {Efros}, \citenamefont {Yugova}, \citenamefont {Oulton}, \citenamefont
  {Stavarache}, \citenamefont {Reuter}, \citenamefont {Wieck},\ and\
  \citenamefont {Bayer}}]{Greilich2006}%
  \BibitemOpen
  \bibfield  {author} {\bibinfo {author} {\bibfnamefont {A.}~\bibnamefont
  {Greilich}}, \bibinfo {author} {\bibfnamefont {D.~R.}\ \bibnamefont
  {Yakovlev}}, \bibinfo {author} {\bibfnamefont {A.}~\bibnamefont {Shabaev}},
  \bibinfo {author} {\bibfnamefont {A.~L.}\ \bibnamefont {Efros}}, \bibinfo
  {author} {\bibfnamefont {I.~A.}\ \bibnamefont {Yugova}}, \bibinfo {author}
  {\bibfnamefont {R.}~\bibnamefont {Oulton}}, \bibinfo {author} {\bibfnamefont
  {V.}~\bibnamefont {Stavarache}}, \bibinfo {author} {\bibfnamefont
  {D.}~\bibnamefont {Reuter}}, \bibinfo {author} {\bibfnamefont
  {A.}~\bibnamefont {Wieck}},\ and\ \bibinfo {author} {\bibfnamefont
  {M.}~\bibnamefont {Bayer}},\ }\bibfield  {title} {\bibinfo {title} {{Mode
  locking of electron spin coherences in singly charged quantum dots}},\ }\href
  {https://doi.org/10.1126/science.1128215} {\bibfield  {journal} {\bibinfo
  {journal} {Science}\ }\textbf {\bibinfo {volume} {313}},\ \bibinfo {pages}
  {341} (\bibinfo {year} {2006})}\BibitemShut {NoStop}%
\bibitem [{\citenamefont {Geim}\ and\ \citenamefont
  {Grigorieva}(2013)}]{Geim2013}%
  \BibitemOpen
  \bibfield  {author} {\bibinfo {author} {\bibfnamefont {A.~K.}\ \bibnamefont
  {Geim}}\ and\ \bibinfo {author} {\bibfnamefont {I.~V.}\ \bibnamefont
  {Grigorieva}},\ }\bibfield  {title} {\bibinfo {title} {{Van der Waals
  heterostructures}},\ }\href {https://doi.org/10.1038/nature12385} {\bibfield
  {journal} {\bibinfo  {journal} {Nature}\ }\textbf {\bibinfo {volume} {499}},\
  \bibinfo {pages} {419} (\bibinfo {year} {2013})},\ \Eprint
  {https://arxiv.org/abs/1307.6718} {arXiv:1307.6718} \BibitemShut {NoStop}%
\bibitem [{\citenamefont {Li}\ \emph {et~al.}(2016)\citenamefont {Li},
  \citenamefont {Qin}, \citenamefont {Xu}, \citenamefont {Cao}, \citenamefont
  {Sun}, \citenamefont {Ma}, \citenamefont {Hu}, \citenamefont {Ren},\ and\
  \citenamefont {Zhen}}]{LiGr}%
  \BibitemOpen
  \bibfield  {author} {\bibinfo {author} {\bibfnamefont {Y.}~\bibnamefont
  {Li}}, \bibinfo {author} {\bibfnamefont {J.-K.}\ \bibnamefont {Qin}},
  \bibinfo {author} {\bibfnamefont {C.-Y.}\ \bibnamefont {Xu}}, \bibinfo
  {author} {\bibfnamefont {J.}~\bibnamefont {Cao}}, \bibinfo {author}
  {\bibfnamefont {Z.-Y.}\ \bibnamefont {Sun}}, \bibinfo {author} {\bibfnamefont
  {L.-P.}\ \bibnamefont {Ma}}, \bibinfo {author} {\bibfnamefont {P.~A.}\
  \bibnamefont {Hu}}, \bibinfo {author} {\bibfnamefont {W.}~\bibnamefont
  {Ren}},\ and\ \bibinfo {author} {\bibfnamefont {L.}~\bibnamefont {Zhen}},\
  }\bibfield  {title} {\bibinfo {title} {{Electric Field Tunable Interlayer
  Relaxation Process and Interlayer Coupling in WSe2/Graphene
  Heterostructures}},\ }\href@noop {} {\bibfield  {journal} {\bibinfo
  {journal} {Advanced Functional Materials}\ }\textbf {\bibinfo {volume}
  {26}},\ \bibinfo {pages} {4319} (\bibinfo {year} {2016})}\BibitemShut
  {NoStop}%
\bibitem [{\citenamefont {Zeng}\ \emph {et~al.}(2012)\citenamefont {Zeng},
  \citenamefont {Dai}, \citenamefont {Yao}, \citenamefont {Xiao},\ and\
  \citenamefont {Cui}}]{Zeng2012}%
  \BibitemOpen
  \bibfield  {author} {\bibinfo {author} {\bibfnamefont {H.}~\bibnamefont
  {Zeng}}, \bibinfo {author} {\bibfnamefont {J.}~\bibnamefont {Dai}}, \bibinfo
  {author} {\bibfnamefont {W.}~\bibnamefont {Yao}}, \bibinfo {author}
  {\bibfnamefont {D.}~\bibnamefont {Xiao}},\ and\ \bibinfo {author}
  {\bibfnamefont {X.}~\bibnamefont {Cui}},\ }\bibfield  {title} {\bibinfo
  {title} {{Valley polarization in MoS$_2$ monolayers by optical pumping}},\
  }\href {https://doi.org/10.1038/nnano.2012.95} {\bibfield  {journal}
  {\bibinfo  {journal} {Nature Nanotechnology}\ }\textbf {\bibinfo {volume}
  {7}},\ \bibinfo {pages} {490} (\bibinfo {year} {2012})},\ \Eprint
  {https://arxiv.org/abs/1202.1592} {arXiv:1202.1592} \BibitemShut {NoStop}%
\bibitem [{\citenamefont {Mak}\ \emph {et~al.}(2012)\citenamefont {Mak},
  \citenamefont {He}, \citenamefont {Shan},\ and\ \citenamefont
  {Heinz}}]{Mak2012}%
  \BibitemOpen
  \bibfield  {author} {\bibinfo {author} {\bibfnamefont {K.~F.}\ \bibnamefont
  {Mak}}, \bibinfo {author} {\bibfnamefont {K.}~\bibnamefont {He}}, \bibinfo
  {author} {\bibfnamefont {J.}~\bibnamefont {Shan}},\ and\ \bibinfo {author}
  {\bibfnamefont {T.~F.}\ \bibnamefont {Heinz}},\ }\bibfield  {title} {\bibinfo
  {title} {{Control of valley polarization in monolayer MoS$_2$ by optical
  helicity}},\ }\href {https://doi.org/10.1038/nnano.2012.96} {\bibfield
  {journal} {\bibinfo  {journal} {Nature Nanotechnology}\ }\textbf {\bibinfo
  {volume} {7}},\ \bibinfo {pages} {494} (\bibinfo {year} {2012})},\ \Eprint
  {https://arxiv.org/abs/1205.1822} {1205.1822} \BibitemShut {NoStop}%
\bibitem [{\citenamefont {Ye}\ \emph {et~al.}(2017)\citenamefont {Ye},
  \citenamefont {Sun},\ and\ \citenamefont {Heinz}}]{Ye2017}%
  \BibitemOpen
  \bibfield  {author} {\bibinfo {author} {\bibfnamefont {Z.}~\bibnamefont
  {Ye}}, \bibinfo {author} {\bibfnamefont {D.}~\bibnamefont {Sun}},\ and\
  \bibinfo {author} {\bibfnamefont {T.}~\bibnamefont {Heinz}},\ }\bibfield
  {title} {\bibinfo {title} {{Optical manipulation of valley pseudospin}},\
  }\href {https://doi.org/10.1038/nphys3891} {\bibfield  {journal} {\bibinfo
  {journal} {Nature Physics}\ }\textbf {\bibinfo {volume} {13}},\ \bibinfo
  {pages} {26} (\bibinfo {year} {2017})}\BibitemShut {NoStop}%
\bibitem [{\citenamefont {Fuhrer}\ and\ \citenamefont
  {Hone}(2013)}]{Fuhrer2013}%
  \BibitemOpen
  \bibfield  {author} {\bibinfo {author} {\bibfnamefont {M.~S.}\ \bibnamefont
  {Fuhrer}}\ and\ \bibinfo {author} {\bibfnamefont {J.}~\bibnamefont {Hone}},\
  }\bibfield  {title} {\bibinfo {title} {{Measurement of mobility in dual-gated
  MoS2 transistors}},\ }\href {https://doi.org/10.1038/nnano.2013.30}
  {\bibfield  {journal} {\bibinfo  {journal} {Nature Nanotechnology}\ }\textbf
  {\bibinfo {volume} {8}},\ \bibinfo {pages} {146} (\bibinfo {year}
  {2013})}\BibitemShut {NoStop}%
\bibitem [{\citenamefont {Mnatsakanov}\ \emph {et~al.}(2004)\citenamefont
  {Mnatsakanov}, \citenamefont {Levinshtein}, \citenamefont {Pomortseva},\ and\
  \citenamefont {Yurkov}}]{Mnatsakanov2004}%
  \BibitemOpen
  \bibfield  {author} {\bibinfo {author} {\bibfnamefont {T.~T.}\ \bibnamefont
  {Mnatsakanov}}, \bibinfo {author} {\bibfnamefont {M.~E.}\ \bibnamefont
  {Levinshtein}}, \bibinfo {author} {\bibfnamefont {L.~I.}\ \bibnamefont
  {Pomortseva}},\ and\ \bibinfo {author} {\bibfnamefont {S.~N.}\ \bibnamefont
  {Yurkov}},\ }\bibfield  {title} {\bibinfo {title} {{Universal Analytical
  Approximation of the Carrier Mobility in Semiconductors for a Wide Range of
  Temperatures and Doping Densities}},\ }\href
  {https://doi.org/10.1134/1.1641133} {\bibfield  {journal} {\bibinfo
  {journal} {Semiconductors}\ }\textbf {\bibinfo {volume} {38}},\ \bibinfo
  {pages} {56} (\bibinfo {year} {2004})}\BibitemShut {NoStop}%
\bibitem [{\citenamefont {Mitioglu}\ \emph {et~al.}(2015)\citenamefont
  {Mitioglu}, \citenamefont {Plochocka}, \citenamefont {{Granados Del Aguila}},
  \citenamefont {Christianen}, \citenamefont {Deligeorgis}, \citenamefont
  {Anghel}, \citenamefont {Kulyuk},\ and\ \citenamefont
  {Maude}}]{Mitioglu2015}%
  \BibitemOpen
  \bibfield  {author} {\bibinfo {author} {\bibfnamefont {A.~A.}\ \bibnamefont
  {Mitioglu}}, \bibinfo {author} {\bibfnamefont {P.}~\bibnamefont {Plochocka}},
  \bibinfo {author} {\bibnamefont {{Granados Del Aguila}}}, \bibinfo {author}
  {\bibfnamefont {P.~C.}\ \bibnamefont {Christianen}}, \bibinfo {author}
  {\bibfnamefont {G.}~\bibnamefont {Deligeorgis}}, \bibinfo {author}
  {\bibfnamefont {S.}~\bibnamefont {Anghel}}, \bibinfo {author} {\bibfnamefont
  {L.}~\bibnamefont {Kulyuk}},\ and\ \bibinfo {author} {\bibfnamefont {D.~K.}\
  \bibnamefont {Maude}},\ }\bibfield  {title} {\bibinfo {title} {{Optical
  Investigation of Monolayer and Bulk Tungsten Diselenide (WSe$_2$) in High
  Magnetic Fields}},\ }\href {https://doi.org/10.1021/acs.nanolett.5b00626}
  {\bibfield  {journal} {\bibinfo  {journal} {Nano Letters}\ }\textbf {\bibinfo
  {volume} {15}},\ \bibinfo {pages} {4387} (\bibinfo {year} {2015})},\ \Eprint
  {https://arxiv.org/abs/1506.03905} {arXiv:1506.03905} \BibitemShut {NoStop}%
\bibitem [{\citenamefont {Plechinger}\ \emph {et~al.}(2016)\citenamefont
  {Plechinger}, \citenamefont {Nagler}, \citenamefont {Arora}, \citenamefont
  {{Granados Del {\'{A}}guila}}, \citenamefont {Ballottin}, \citenamefont
  {Frank}, \citenamefont {Steinleitner}, \citenamefont {Gmitra}, \citenamefont
  {Fabian}, \citenamefont {Christianen}, \citenamefont {Bratschitsch},
  \citenamefont {Sch{\"{u}}ller},\ and\ \citenamefont {Korn}}]{Plechinger2016}%
  \BibitemOpen
  \bibfield  {author} {\bibinfo {author} {\bibfnamefont {G.}~\bibnamefont
  {Plechinger}}, \bibinfo {author} {\bibfnamefont {P.}~\bibnamefont {Nagler}},
  \bibinfo {author} {\bibfnamefont {A.}~\bibnamefont {Arora}}, \bibinfo
  {author} {\bibfnamefont {A.}~\bibnamefont {{Granados Del {\'{A}}guila}}},
  \bibinfo {author} {\bibfnamefont {M.~V.}\ \bibnamefont {Ballottin}}, \bibinfo
  {author} {\bibfnamefont {T.}~\bibnamefont {Frank}}, \bibinfo {author}
  {\bibfnamefont {P.}~\bibnamefont {Steinleitner}}, \bibinfo {author}
  {\bibfnamefont {M.}~\bibnamefont {Gmitra}}, \bibinfo {author} {\bibfnamefont
  {J.}~\bibnamefont {Fabian}}, \bibinfo {author} {\bibfnamefont {P.~C.}\
  \bibnamefont {Christianen}}, \bibinfo {author} {\bibfnamefont
  {R.}~\bibnamefont {Bratschitsch}}, \bibinfo {author} {\bibfnamefont
  {C.}~\bibnamefont {Sch{\"{u}}ller}},\ and\ \bibinfo {author} {\bibfnamefont
  {T.}~\bibnamefont {Korn}},\ }\bibfield  {title} {\bibinfo {title} {{Excitonic
  Valley Effects in Monolayer WS$_2$ under High Magnetic Fields}},\ }\href
  {https://doi.org/10.1021/acs.nanolett.6b04171} {\bibfield  {journal}
  {\bibinfo  {journal} {Nano Letters}\ }\textbf {\bibinfo {volume} {16}},\
  \bibinfo {pages} {7899} (\bibinfo {year} {2016})},\ \Eprint
  {https://arxiv.org/abs/1612.03004} {arXiv:1612.03004} \BibitemShut {NoStop}%
\bibitem [{\citenamefont {Li}\ and\ \citenamefont {Appelbaum}(2015)}]{Li2015}%
  \BibitemOpen
  \bibfield  {author} {\bibinfo {author} {\bibfnamefont {P.}~\bibnamefont
  {Li}}\ and\ \bibinfo {author} {\bibfnamefont {I.}~\bibnamefont {Appelbaum}},\
  }\bibfield  {title} {\bibinfo {title} {{Symmetry, distorted band structure,
  and spin-orbit coupling of group-III metal-monochalcogenide monolayers}},\
  }\href {https://doi.org/10.1103/PhysRevB.92.195129} {\bibfield  {journal}
  {\bibinfo  {journal} {Phys. Rev. B}\ }\textbf {\bibinfo {volume} {92}},\
  \bibinfo {pages} {195129} (\bibinfo {year} {2015})}\BibitemShut {NoStop}%
\bibitem [{\citenamefont {Do}\ \emph {et~al.}(2015)\citenamefont {Do},
  \citenamefont {Mahanti},\ and\ \citenamefont {Lai}}]{Do2015}%
  \BibitemOpen
  \bibfield  {author} {\bibinfo {author} {\bibfnamefont {D.~T.}\ \bibnamefont
  {Do}}, \bibinfo {author} {\bibfnamefont {S.~D.}\ \bibnamefont {Mahanti}},\
  and\ \bibinfo {author} {\bibfnamefont {C.~W.}\ \bibnamefont {Lai}},\
  }\bibfield  {title} {\bibinfo {title} {{Spin splitting in 2D monochalcogenide
  semiconductors}},\ }\href {https://doi.org/10.1038/srep17044} {\bibfield
  {journal} {\bibinfo  {journal} {Scientific Reports}\ }\textbf {\bibinfo
  {volume} {5}},\ \bibinfo {pages} {17044} (\bibinfo {year}
  {2015})}\BibitemShut {NoStop}%
\bibitem [{\citenamefont {Premasiri}\ and\ \citenamefont
  {Gao}(2019)}]{Premasiri2019}%
  \BibitemOpen
  \bibfield  {author} {\bibinfo {author} {\bibfnamefont {K.}~\bibnamefont
  {Premasiri}}\ and\ \bibinfo {author} {\bibfnamefont {X.~P.}\ \bibnamefont
  {Gao}},\ }\bibfield  {title} {\bibinfo {title} {{Tuning spin-orbit coupling
  in 2D materials for spintronics: A topical review}},\ }\href@noop {}
  {\bibfield  {journal} {\bibinfo  {journal} {Journal of Physics Condensed
  Matter}\ }\textbf {\bibinfo {volume} {31}} (\bibinfo {year}
  {2019})}\BibitemShut {NoStop}%
\bibitem [{\citenamefont {Song}\ \emph {et~al.}(2020)\citenamefont {Song},
  \citenamefont {Huang}, \citenamefont {Wang}, \citenamefont {Luo},\ and\
  \citenamefont {Yan}}]{Song2020}%
  \BibitemOpen
  \bibfield  {author} {\bibinfo {author} {\bibfnamefont {C.}~\bibnamefont
  {Song}}, \bibinfo {author} {\bibfnamefont {S.}~\bibnamefont {Huang}},
  \bibinfo {author} {\bibfnamefont {C.}~\bibnamefont {Wang}}, \bibinfo {author}
  {\bibfnamefont {J.}~\bibnamefont {Luo}},\ and\ \bibinfo {author}
  {\bibfnamefont {H.}~\bibnamefont {Yan}},\ }\bibfield  {title} {\bibinfo
  {title} {{The optical properties of few-layer InSe}},\ }\bibfield  {journal}
  {\bibinfo  {journal} {Journal of Applied Physics}\ }\textbf {\bibinfo
  {volume} {128}},\ \href {https://doi.org/10.1063/5.0018480}
  {10.1063/5.0018480} (\bibinfo {year} {2020}),\ \Eprint
  {https://arxiv.org/abs/2008.04691} {arXiv:2008.04691} \BibitemShut {NoStop}%
\bibitem [{\citenamefont {Bandurin}\ \emph {et~al.}(2017)\citenamefont
  {Bandurin}, \citenamefont {Tyurnina}, \citenamefont {Yu}, \citenamefont
  {Mishchenko}, \citenamefont {Z{\'{o}}lyomi}, \citenamefont {Morozov},
  \citenamefont {Kumar}, \citenamefont {Gorbachev}, \citenamefont {Kudrynskyi},
  \citenamefont {Pezzini}, \citenamefont {Kovalyuk}, \citenamefont {Zeitler},
  \citenamefont {Novoselov}, \citenamefont {Patan{\`{e}}}, \citenamefont
  {Eaves}, \citenamefont {Grigorieva}, \citenamefont {Fal'ko}, \citenamefont
  {Geim},\ and\ \citenamefont {Cao}}]{Bandurin2017}%
  \BibitemOpen
  \bibfield  {author} {\bibinfo {author} {\bibfnamefont {D.~A.}\ \bibnamefont
  {Bandurin}}, \bibinfo {author} {\bibfnamefont {A.~V.}\ \bibnamefont
  {Tyurnina}}, \bibinfo {author} {\bibfnamefont {G.~L.}\ \bibnamefont {Yu}},
  \bibinfo {author} {\bibfnamefont {A.}~\bibnamefont {Mishchenko}}, \bibinfo
  {author} {\bibfnamefont {V.}~\bibnamefont {Z{\'{o}}lyomi}}, \bibinfo {author}
  {\bibfnamefont {S.~V.}\ \bibnamefont {Morozov}}, \bibinfo {author}
  {\bibfnamefont {R.~K.}\ \bibnamefont {Kumar}}, \bibinfo {author}
  {\bibfnamefont {R.~V.}\ \bibnamefont {Gorbachev}}, \bibinfo {author}
  {\bibfnamefont {Z.~R.}\ \bibnamefont {Kudrynskyi}}, \bibinfo {author}
  {\bibfnamefont {S.}~\bibnamefont {Pezzini}}, \bibinfo {author} {\bibfnamefont
  {Z.~D.}\ \bibnamefont {Kovalyuk}}, \bibinfo {author} {\bibfnamefont
  {U.}~\bibnamefont {Zeitler}}, \bibinfo {author} {\bibfnamefont {K.~S.}\
  \bibnamefont {Novoselov}}, \bibinfo {author} {\bibfnamefont {A.}~\bibnamefont
  {Patan{\`{e}}}}, \bibinfo {author} {\bibfnamefont {L.}~\bibnamefont {Eaves}},
  \bibinfo {author} {\bibfnamefont {I.~V.}\ \bibnamefont {Grigorieva}},
  \bibinfo {author} {\bibfnamefont {V.~I.}\ \bibnamefont {Fal'ko}}, \bibinfo
  {author} {\bibfnamefont {A.~K.}\ \bibnamefont {Geim}},\ and\ \bibinfo
  {author} {\bibfnamefont {Y.}~\bibnamefont {Cao}},\ }\bibfield  {title}
  {\bibinfo {title} {{High electron mobility, quantum Hall effect and anomalous
  optical response in atomically thin InSe}},\ }\href
  {https://doi.org/10.1038/nnano.2016.242} {\bibfield  {journal} {\bibinfo
  {journal} {Nature Nanotechnology}\ }\textbf {\bibinfo {volume} {12}},\
  \bibinfo {pages} {223} (\bibinfo {year} {2017})}\BibitemShut {NoStop}%
\bibitem [{\citenamefont {Venanzi}\ \emph {et~al.}(2020)\citenamefont
  {Venanzi}, \citenamefont {Arora}, \citenamefont {Winnerl}, \citenamefont
  {Pashkin}, \citenamefont {Chava}, \citenamefont {Patan\`e}, \citenamefont
  {Kovalyuk}, \citenamefont {Kudrynskyi}, \citenamefont {Watanabe},
  \citenamefont {Taniguchi}, \citenamefont {Erbe}, \citenamefont {Helm},\ and\
  \citenamefont {Schneider}}]{Venanzi2020}%
  \BibitemOpen
  \bibfield  {author} {\bibinfo {author} {\bibfnamefont {T.}~\bibnamefont
  {Venanzi}}, \bibinfo {author} {\bibfnamefont {H.}~\bibnamefont {Arora}},
  \bibinfo {author} {\bibfnamefont {S.}~\bibnamefont {Winnerl}}, \bibinfo
  {author} {\bibfnamefont {A.}~\bibnamefont {Pashkin}}, \bibinfo {author}
  {\bibfnamefont {P.}~\bibnamefont {Chava}}, \bibinfo {author} {\bibfnamefont
  {A.}~\bibnamefont {Patan\`e}}, \bibinfo {author} {\bibfnamefont {Z.~D.}\
  \bibnamefont {Kovalyuk}}, \bibinfo {author} {\bibfnamefont {Z.~R.}\
  \bibnamefont {Kudrynskyi}}, \bibinfo {author} {\bibfnamefont
  {K.}~\bibnamefont {Watanabe}}, \bibinfo {author} {\bibfnamefont
  {T.}~\bibnamefont {Taniguchi}}, \bibinfo {author} {\bibfnamefont
  {A.}~\bibnamefont {Erbe}}, \bibinfo {author} {\bibfnamefont {M.}~\bibnamefont
  {Helm}},\ and\ \bibinfo {author} {\bibfnamefont {H.}~\bibnamefont
  {Schneider}},\ }\bibfield  {title} {\bibinfo {title} {Photoluminescence
  dynamics in few-layer inse},\ }\href
  {https://doi.org/10.1103/PhysRevMaterials.4.044001} {\bibfield  {journal}
  {\bibinfo  {journal} {Phys. Rev. Materials}\ }\textbf {\bibinfo {volume}
  {4}},\ \bibinfo {pages} {044001} (\bibinfo {year} {2020})}\BibitemShut
  {NoStop}%
\bibitem [{\citenamefont {Shubina}\ \emph {et~al.}(2019)\citenamefont
  {Shubina}, \citenamefont {Desrat}, \citenamefont {Moret}, \citenamefont
  {Tiberj}, \citenamefont {Briot}, \citenamefont {Davydov}, \citenamefont
  {Platonov}, \citenamefont {Semina},\ and\ \citenamefont {Gil}}]{Shubina2019}%
  \BibitemOpen
  \bibfield  {author} {\bibinfo {author} {\bibfnamefont {T.~V.}\ \bibnamefont
  {Shubina}}, \bibinfo {author} {\bibfnamefont {W.}~\bibnamefont {Desrat}},
  \bibinfo {author} {\bibfnamefont {M.}~\bibnamefont {Moret}}, \bibinfo
  {author} {\bibfnamefont {A.}~\bibnamefont {Tiberj}}, \bibinfo {author}
  {\bibfnamefont {O.}~\bibnamefont {Briot}}, \bibinfo {author} {\bibfnamefont
  {V.~Y.}\ \bibnamefont {Davydov}}, \bibinfo {author} {\bibfnamefont {A.~V.}\
  \bibnamefont {Platonov}}, \bibinfo {author} {\bibfnamefont {M.~A.}\
  \bibnamefont {Semina}},\ and\ \bibinfo {author} {\bibfnamefont
  {B.}~\bibnamefont {Gil}},\ }\bibfield  {title} {\bibinfo {title} {{InSe as a
  case between 3D and 2D layered crystals for excitons}},\ }\href
  {https://doi.org/10.1038/s41467-019-11487-0} {\bibfield  {journal} {\bibinfo
  {journal} {Nature Communications}\ }\textbf {\bibinfo {volume} {10}},\
  \bibinfo {pages} {1} (\bibinfo {year} {2019})},\ \Eprint
  {https://arxiv.org/abs/1904.00390} {arXiv:1904.00390} \BibitemShut {NoStop}%
\bibitem [{\citenamefont {Magorrian}\ \emph {et~al.}(2016)\citenamefont
  {Magorrian}, \citenamefont {Z\'olyomi},\ and\ \citenamefont
  {Fal'ko}}]{Magorrian2016}%
  \BibitemOpen
  \bibfield  {author} {\bibinfo {author} {\bibfnamefont {S.~J.}\ \bibnamefont
  {Magorrian}}, \bibinfo {author} {\bibfnamefont {V.}~\bibnamefont
  {Z\'olyomi}},\ and\ \bibinfo {author} {\bibfnamefont {V.~I.}\ \bibnamefont
  {Fal'ko}},\ }\bibfield  {title} {\bibinfo {title} {{Electronic and optical
  properties of two-dimensional InSe from a DFT-parametrized tight-binding
  model}},\ }\href {https://doi.org/10.1103/PhysRevB.94.245431} {\bibfield
  {journal} {\bibinfo  {journal} {Phys. Rev. B}\ }\textbf {\bibinfo {volume}
  {94}},\ \bibinfo {pages} {245431} (\bibinfo {year} {2016})}\BibitemShut
  {NoStop}%
\bibitem [{\citenamefont {Magorrian}\ \emph {et~al.}(2017)\citenamefont
  {Magorrian}, \citenamefont {Z{\'{o}}lyomi},\ and\ \citenamefont
  {Fal'ko}}]{Magorrian2017}%
  \BibitemOpen
  \bibfield  {author} {\bibinfo {author} {\bibfnamefont {S.~J.}\ \bibnamefont
  {Magorrian}}, \bibinfo {author} {\bibfnamefont {V.}~\bibnamefont
  {Z{\'{o}}lyomi}},\ and\ \bibinfo {author} {\bibfnamefont {V.~I.}\
  \bibnamefont {Fal'ko}},\ }\bibfield  {title} {\bibinfo {title} {{Spin-orbit
  coupling, optical transitions, and spin pumping in monolayer and few-layer
  InSe}},\ }\href {https://doi.org/10.1103/PhysRevB.96.195428} {\bibfield
  {journal} {\bibinfo  {journal} {Physical Review B}\ }\textbf {\bibinfo
  {volume} {96}},\ \bibinfo {pages} {195428} (\bibinfo {year} {2017})},\
  \Eprint {https://arxiv.org/abs/1711.03402} {arXiv:1711.03402} \BibitemShut
  {NoStop}%
\bibitem [{\citenamefont {Dey}\ \emph {et~al.}(2017)\citenamefont {Dey},
  \citenamefont {Yang}, \citenamefont {Robert}, \citenamefont {Wang},
  \citenamefont {Urbaszek}, \citenamefont {Marie},\ and\ \citenamefont
  {Crooker}}]{Dey2017}%
  \BibitemOpen
  \bibfield  {author} {\bibinfo {author} {\bibfnamefont {P.}~\bibnamefont
  {Dey}}, \bibinfo {author} {\bibfnamefont {L.}~\bibnamefont {Yang}}, \bibinfo
  {author} {\bibfnamefont {C.}~\bibnamefont {Robert}}, \bibinfo {author}
  {\bibfnamefont {G.}~\bibnamefont {Wang}}, \bibinfo {author} {\bibfnamefont
  {B.}~\bibnamefont {Urbaszek}}, \bibinfo {author} {\bibfnamefont
  {X.}~\bibnamefont {Marie}},\ and\ \bibinfo {author} {\bibfnamefont {S.~A.}\
  \bibnamefont {Crooker}},\ }\bibfield  {title} {\bibinfo {title}
  {{Gate-Controlled Spin-Valley Locking of Resident Carriers in WSe$_2$
  Monolayers}},\ }\href {https://doi.org/10.1103/PhysRevLett.119.137401}
  {\bibfield  {journal} {\bibinfo  {journal} {Physical Review Letters}\
  }\textbf {\bibinfo {volume} {119}},\ \bibinfo {pages} {137401} (\bibinfo
  {year} {2017})},\ \Eprint {https://arxiv.org/abs/1704.05448}
  {arXiv:1704.05448} \BibitemShut {NoStop}%
\bibitem [{\citenamefont {Tang}\ \emph {et~al.}(2015)\citenamefont {Tang},
  \citenamefont {Xie}, \citenamefont {Mandal}, \citenamefont {McGuire},\ and\
  \citenamefont {Lai}}]{Tang2015}%
  \BibitemOpen
  \bibfield  {author} {\bibinfo {author} {\bibfnamefont {Y.}~\bibnamefont
  {Tang}}, \bibinfo {author} {\bibfnamefont {W.}~\bibnamefont {Xie}}, \bibinfo
  {author} {\bibfnamefont {K.~C.}\ \bibnamefont {Mandal}}, \bibinfo {author}
  {\bibfnamefont {J.~A.}\ \bibnamefont {McGuire}},\ and\ \bibinfo {author}
  {\bibfnamefont {C.~W.}\ \bibnamefont {Lai}},\ }\bibfield  {title} {\bibinfo
  {title} {{Optical and spin polarization dynamics in GaSe nanoslabs}},\ }\href
  {https://doi.org/10.1103/PhysRevB.91.195429} {\bibfield  {journal} {\bibinfo
  {journal} {Phys. Rev. B}\ }\textbf {\bibinfo {volume} {91}},\ \bibinfo
  {pages} {195429} (\bibinfo {year} {2015})}\BibitemShut {NoStop}%
\bibitem [{\citenamefont {Gamarts}\ \emph {et~al.}(1977)\citenamefont
  {Gamarts}, \citenamefont {Ivchenko}, \citenamefont {Karaman}, \citenamefont
  {Mushinskiǐ}, \citenamefont {Pikus}, \citenamefont {Razbirin},\ and\
  \citenamefont {Starukhin}}]{Gamarts1977a}%
  \BibitemOpen
  \bibfield  {author} {\bibinfo {author} {\bibfnamefont {E.}~\bibnamefont
  {Gamarts}}, \bibinfo {author} {\bibfnamefont {E.}~\bibnamefont {Ivchenko}},
  \bibinfo {author} {\bibfnamefont {M.}~\bibnamefont {Karaman}}, \bibinfo
  {author} {\bibfnamefont {V.}~\bibnamefont {Mushinskiǐ}}, \bibinfo {author}
  {\bibfnamefont {G.}~\bibnamefont {Pikus}}, \bibinfo {author} {\bibfnamefont
  {B.}~\bibnamefont {Razbirin}},\ and\ \bibinfo {author} {\bibfnamefont
  {A.}~\bibnamefont {Starukhin}},\ }\bibfield  {title} {\bibinfo {title}
  {{Optical orientation and alignment of free excitons in GaSe during resonance
  excitation. Experiment}},\ }\href@noop {} {\bibfield  {journal} {\bibinfo
  {journal} {Soviet Journal of Experimental and Theoretical Physics}\ }\textbf
  {\bibinfo {volume} {46}},\ \bibinfo {pages} {590} (\bibinfo {year}
  {1977})}\BibitemShut {NoStop}%
\bibitem [{\citenamefont {Arora}\ and\ \citenamefont {Erbe}(2021)}]{Arora2021}%
  \BibitemOpen
  \bibfield  {author} {\bibinfo {author} {\bibfnamefont {H.}~\bibnamefont
  {Arora}}\ and\ \bibinfo {author} {\bibfnamefont {A.}~\bibnamefont {Erbe}},\
  }\bibfield  {title} {\bibinfo {title} {{Recent progress in contact, mobility,
  and encapsulation engineering of InSe and GaSe}},\ }\href
  {https://doi.org/10.1002/inf2.12160} {\bibfield  {journal} {\bibinfo
  {journal} {InfoMat}\ }\textbf {\bibinfo {volume} {3}},\ \bibinfo {pages}
  {662} (\bibinfo {year} {2021})}\BibitemShut {NoStop}%
\bibitem [{\citenamefont {Late}\ \emph {et~al.}(2012)\citenamefont {Late},
  \citenamefont {Liu}, \citenamefont {Luo}, \citenamefont {Yan}, \citenamefont
  {Matte}, \citenamefont {Grayson}, \citenamefont {Rao},\ and\ \citenamefont
  {Dravid}}]{Late2012}%
  \BibitemOpen
  \bibfield  {author} {\bibinfo {author} {\bibfnamefont {D.~J.}\ \bibnamefont
  {Late}}, \bibinfo {author} {\bibfnamefont {B.}~\bibnamefont {Liu}}, \bibinfo
  {author} {\bibfnamefont {J.}~\bibnamefont {Luo}}, \bibinfo {author}
  {\bibfnamefont {A.}~\bibnamefont {Yan}}, \bibinfo {author} {\bibfnamefont
  {H.~S. S.~R.}\ \bibnamefont {Matte}}, \bibinfo {author} {\bibfnamefont
  {M.}~\bibnamefont {Grayson}}, \bibinfo {author} {\bibfnamefont {C.~N.~R.}\
  \bibnamefont {Rao}},\ and\ \bibinfo {author} {\bibfnamefont {V.~P.}\
  \bibnamefont {Dravid}},\ }\bibfield  {title} {\bibinfo {title} {{GaS and GaSe
  Ultrathin Layer Transistors}},\ }\href@noop {} {\bibfield  {journal}
  {\bibinfo  {journal} {Advanced Materials}\ }\textbf {\bibinfo {volume}
  {24}},\ \bibinfo {pages} {3549} (\bibinfo {year} {2012})}\BibitemShut
  {NoStop}%
\bibitem [{\citenamefont {Ivchenko}\ \emph {et~al.}(1977)\citenamefont
  {Ivchenko}, \citenamefont {Pikus}, \citenamefont {Razbirin},\ and\
  \citenamefont {Starukhin}}]{Ivchenko1977}%
  \BibitemOpen
  \bibfield  {author} {\bibinfo {author} {\bibfnamefont {E.}~\bibnamefont
  {Ivchenko}}, \bibinfo {author} {\bibfnamefont {G.}~\bibnamefont {Pikus}},
  \bibinfo {author} {\bibfnamefont {B.}~\bibnamefont {Razbirin}},\ and\
  \bibinfo {author} {\bibfnamefont {A.}~\bibnamefont {Starukhin}},\ }\bibfield
  {title} {\bibinfo {title} {{Optical orientation and alignment of free
  excitons in GaSe during resonance excitation. Theory}},\ }\href@noop {}
  {\bibfield  {journal} {\bibinfo  {journal} {Soviet Journal of Experimental
  and Theoretical Physics}\ }\textbf {\bibinfo {volume} {46}},\ \bibinfo
  {pages} {590} (\bibinfo {year} {1977})}\BibitemShut {NoStop}%
\bibitem [{\citenamefont {Ceferino}\ \emph {et~al.}(2021)\citenamefont
  {Ceferino}, \citenamefont {Magorrian}, \citenamefont {Z\'olyomi},
  \citenamefont {Bandurin}, \citenamefont {Geim}, \citenamefont {Patan\`e},
  \citenamefont {Kovalyuk}, \citenamefont {Kudrynskyi}, \citenamefont
  {Grigorieva},\ and\ \citenamefont {Fal'ko}}]{Ceferino2021}%
  \BibitemOpen
  \bibfield  {author} {\bibinfo {author} {\bibfnamefont {A.}~\bibnamefont
  {Ceferino}}, \bibinfo {author} {\bibfnamefont {S.~J.}\ \bibnamefont
  {Magorrian}}, \bibinfo {author} {\bibfnamefont {V.}~\bibnamefont
  {Z\'olyomi}}, \bibinfo {author} {\bibfnamefont {D.~A.}\ \bibnamefont
  {Bandurin}}, \bibinfo {author} {\bibfnamefont {A.~K.}\ \bibnamefont {Geim}},
  \bibinfo {author} {\bibfnamefont {A.}~\bibnamefont {Patan\`e}}, \bibinfo
  {author} {\bibfnamefont {Z.~D.}\ \bibnamefont {Kovalyuk}}, \bibinfo {author}
  {\bibfnamefont {Z.~R.}\ \bibnamefont {Kudrynskyi}}, \bibinfo {author}
  {\bibfnamefont {I.~V.}\ \bibnamefont {Grigorieva}},\ and\ \bibinfo {author}
  {\bibfnamefont {V.~I.}\ \bibnamefont {Fal'ko}},\ }\bibfield  {title}
  {\bibinfo {title} {{Tunable spin-orbit coupling in two-dimensional InSe}},\
  }\href {https://doi.org/10.1103/PhysRevB.104.125432} {\bibfield  {journal}
  {\bibinfo  {journal} {Phys. Rev. B}\ }\textbf {\bibinfo {volume} {104}},\
  \bibinfo {pages} {125432} (\bibinfo {year} {2021})}\BibitemShut {NoStop}%
\bibitem [{\citenamefont {Kuroda}\ \emph {et~al.}(1980)\citenamefont {Kuroda},
  \citenamefont {Munakata},\ and\ \citenamefont {Nishina}}]{KURODA1980}%
  \BibitemOpen
  \bibfield  {author} {\bibinfo {author} {\bibfnamefont {N.}~\bibnamefont
  {Kuroda}}, \bibinfo {author} {\bibfnamefont {I.}~\bibnamefont {Munakata}},\
  and\ \bibinfo {author} {\bibfnamefont {Y.}~\bibnamefont {Nishina}},\
  }\bibfield  {title} {\bibinfo {title} {{Exciton transitions from spin-orbit
  split off valence bands in layer compound InSe}},\ }\href@noop {} {\bibfield
  {journal} {\bibinfo  {journal} {Solid State Communications}\ }\textbf
  {\bibinfo {volume} {33}},\ \bibinfo {pages} {687} (\bibinfo {year}
  {1980})}\BibitemShut {NoStop}%
\bibitem [{\citenamefont {Brotons-Gisbert}\ \emph {et~al.}(2019)\citenamefont
  {Brotons-Gisbert}, \citenamefont {Proux}, \citenamefont {Picard},
  \citenamefont {Andres-Penares}, \citenamefont {Branny}, \citenamefont
  {Molina-S{\'{a}}nchez}, \citenamefont {S{\'{a}}nchez-Royo},\ and\
  \citenamefont {Gerardot}}]{Brotons-Gisbert2019}%
  \BibitemOpen
  \bibfield  {author} {\bibinfo {author} {\bibfnamefont {M.}~\bibnamefont
  {Brotons-Gisbert}}, \bibinfo {author} {\bibfnamefont {R.}~\bibnamefont
  {Proux}}, \bibinfo {author} {\bibfnamefont {R.}~\bibnamefont {Picard}},
  \bibinfo {author} {\bibfnamefont {D.}~\bibnamefont {Andres-Penares}},
  \bibinfo {author} {\bibfnamefont {A.}~\bibnamefont {Branny}}, \bibinfo
  {author} {\bibfnamefont {A.}~\bibnamefont {Molina-S{\'{a}}nchez}}, \bibinfo
  {author} {\bibfnamefont {J.~F.}\ \bibnamefont {S{\'{a}}nchez-Royo}},\ and\
  \bibinfo {author} {\bibfnamefont {B.~D.}\ \bibnamefont {Gerardot}},\
  }\bibfield  {title} {\bibinfo {title} {{Out-of-plane orientation of
  luminescent excitons in two-dimensional indium selenide}},\ }\href
  {https://doi.org/10.1038/s41467-019-11920-4} {\bibfield  {journal} {\bibinfo
  {journal} {Nature Communications}\ }\textbf {\bibinfo {volume} {10}},\
  \bibinfo {pages} {1} (\bibinfo {year} {2019})}\BibitemShut {NoStop}%
\bibitem [{\citenamefont {Wang}\ \emph {et~al.}(2013)\citenamefont {Wang},
  \citenamefont {Meric}, \citenamefont {Huang}, \citenamefont {Gao},
  \citenamefont {Gao}, \citenamefont {Tran}, \citenamefont {Taniguchi},
  \citenamefont {Watanabe}, \citenamefont {Campos}, \citenamefont {Muller},
  \citenamefont {Guo}, \citenamefont {Kim}, \citenamefont {Hone}, \citenamefont
  {Shepard},\ and\ \citenamefont {Dean}}]{Wang2013}%
  \BibitemOpen
  \bibfield  {author} {\bibinfo {author} {\bibfnamefont {L.}~\bibnamefont
  {Wang}}, \bibinfo {author} {\bibfnamefont {I.}~\bibnamefont {Meric}},
  \bibinfo {author} {\bibfnamefont {P.~Y.}\ \bibnamefont {Huang}}, \bibinfo
  {author} {\bibfnamefont {Q.}~\bibnamefont {Gao}}, \bibinfo {author}
  {\bibfnamefont {Y.}~\bibnamefont {Gao}}, \bibinfo {author} {\bibfnamefont
  {H.}~\bibnamefont {Tran}}, \bibinfo {author} {\bibfnamefont {T.}~\bibnamefont
  {Taniguchi}}, \bibinfo {author} {\bibfnamefont {K.}~\bibnamefont {Watanabe}},
  \bibinfo {author} {\bibfnamefont {L.~M.}\ \bibnamefont {Campos}}, \bibinfo
  {author} {\bibfnamefont {D.~A.}\ \bibnamefont {Muller}}, \bibinfo {author}
  {\bibfnamefont {J.}~\bibnamefont {Guo}}, \bibinfo {author} {\bibfnamefont
  {P.}~\bibnamefont {Kim}}, \bibinfo {author} {\bibfnamefont {J.}~\bibnamefont
  {Hone}}, \bibinfo {author} {\bibfnamefont {K.~L.}\ \bibnamefont {Shepard}},\
  and\ \bibinfo {author} {\bibfnamefont {C.~R.}\ \bibnamefont {Dean}},\
  }\bibfield  {title} {\bibinfo {title} {One-dimensional electrical contact to
  a two-dimensional material},\ }\href
  {https://doi.org/10.1126/science.1244358} {\bibfield  {journal} {\bibinfo
  {journal} {Science}\ }\textbf {\bibinfo {volume} {342}},\ \bibinfo {pages}
  {614} (\bibinfo {year} {2013})}\BibitemShut {NoStop}%
\bibitem [{\citenamefont {Kretinin}\ \emph {et~al.}(2014)\citenamefont
  {Kretinin}, \citenamefont {Cao}, \citenamefont {Tu}, \citenamefont {Yu},
  \citenamefont {Jalil}, \citenamefont {Novoselov}, \citenamefont {Haigh},
  \citenamefont {Gholinia}, \citenamefont {Mishchenko}, \citenamefont {Lozada},
  \citenamefont {Georgiou}, \citenamefont {Woods}, \citenamefont {Withers},
  \citenamefont {Blake}, \citenamefont {Eda}, \citenamefont {Wirsig},
  \citenamefont {Hucho}, \citenamefont {Watanabe}, \citenamefont {Taniguchi},
  \citenamefont {Geim},\ and\ \citenamefont {Gorbachev}}]{Kretinin2014}%
  \BibitemOpen
  \bibfield  {author} {\bibinfo {author} {\bibfnamefont {A.~V.}\ \bibnamefont
  {Kretinin}}, \bibinfo {author} {\bibfnamefont {Y.}~\bibnamefont {Cao}},
  \bibinfo {author} {\bibfnamefont {J.~S.}\ \bibnamefont {Tu}}, \bibinfo
  {author} {\bibfnamefont {G.~L.}\ \bibnamefont {Yu}}, \bibinfo {author}
  {\bibfnamefont {R.}~\bibnamefont {Jalil}}, \bibinfo {author} {\bibfnamefont
  {K.~S.}\ \bibnamefont {Novoselov}}, \bibinfo {author} {\bibfnamefont {S.~J.}\
  \bibnamefont {Haigh}}, \bibinfo {author} {\bibfnamefont {A.}~\bibnamefont
  {Gholinia}}, \bibinfo {author} {\bibfnamefont {A.}~\bibnamefont
  {Mishchenko}}, \bibinfo {author} {\bibfnamefont {M.}~\bibnamefont {Lozada}},
  \bibinfo {author} {\bibfnamefont {T.}~\bibnamefont {Georgiou}}, \bibinfo
  {author} {\bibfnamefont {C.~R.}\ \bibnamefont {Woods}}, \bibinfo {author}
  {\bibfnamefont {F.}~\bibnamefont {Withers}}, \bibinfo {author} {\bibfnamefont
  {P.}~\bibnamefont {Blake}}, \bibinfo {author} {\bibfnamefont
  {G.}~\bibnamefont {Eda}}, \bibinfo {author} {\bibfnamefont {A.}~\bibnamefont
  {Wirsig}}, \bibinfo {author} {\bibfnamefont {C.}~\bibnamefont {Hucho}},
  \bibinfo {author} {\bibfnamefont {K.}~\bibnamefont {Watanabe}}, \bibinfo
  {author} {\bibfnamefont {T.}~\bibnamefont {Taniguchi}}, \bibinfo {author}
  {\bibfnamefont {A.~K.}\ \bibnamefont {Geim}},\ and\ \bibinfo {author}
  {\bibfnamefont {R.~V.}\ \bibnamefont {Gorbachev}},\ }\bibfield  {title}
  {\bibinfo {title} {{Electronic Properties of Graphene Encapsulated with
  Different Two-Dimensional Atomic Crystals}},\ }\href@noop {} {\bibfield
  {journal} {\bibinfo  {journal} {Nano Letters}\ }\textbf {\bibinfo {volume}
  {14}},\ \bibinfo {pages} {3270} (\bibinfo {year} {2014})}\BibitemShut
  {NoStop}%
\bibitem [{\citenamefont {Zhong}\ \emph {et~al.}(2019)\citenamefont {Zhong},
  \citenamefont {Sangwan}, \citenamefont {Kang}, \citenamefont {Luxa},
  \citenamefont {Hersam},\ and\ \citenamefont {Weiss}}]{Zhong2019}%
  \BibitemOpen
  \bibfield  {author} {\bibinfo {author} {\bibfnamefont {C.}~\bibnamefont
  {Zhong}}, \bibinfo {author} {\bibfnamefont {V.~K.}\ \bibnamefont {Sangwan}},
  \bibinfo {author} {\bibfnamefont {J.}~\bibnamefont {Kang}}, \bibinfo {author}
  {\bibfnamefont {J.}~\bibnamefont {Luxa}}, \bibinfo {author} {\bibfnamefont
  {M.~C.}\ \bibnamefont {Hersam}},\ and\ \bibinfo {author} {\bibfnamefont
  {E.~A.}\ \bibnamefont {Weiss}},\ }\bibfield  {title} {\bibinfo {title} {{Hot
  Carrier and Surface Recombination Dynamics in Layered InSe Crystals}},\
  }\href {https://doi.org/10.1021/acs.jpclett.8b03543} {\bibfield  {journal}
  {\bibinfo  {journal} {J. Phys. Chem. Lett}\ }\textbf {\bibinfo {volume}
  {10}},\ \bibinfo {pages} {499} (\bibinfo {year} {2019})}\BibitemShut
  {NoStop}%
\bibitem [{\citenamefont {Kuroda}\ \emph {et~al.}(1982)\citenamefont {Kuroda},
  \citenamefont {Munakata},\ and\ \citenamefont {Nishina}}]{Kuroda1982}%
  \BibitemOpen
  \bibfield  {author} {\bibinfo {author} {\bibfnamefont {N.}~\bibnamefont
  {Kuroda}}, \bibinfo {author} {\bibfnamefont {I.}~\bibnamefont {Munakata}},\
  and\ \bibinfo {author} {\bibfnamefont {Y.}~\bibnamefont {Nishina}},\
  }\bibfield  {title} {\bibinfo {title} {{Exciton Selection Rules in the
  Polarized Resonant Raman Scattering by LO Phonons in InSe}},\ }\href
  {https://doi.org/10.1143/JPSJ.51.839} {\bibfield  {journal} {\bibinfo
  {journal} {Journal of the Physical Society of Japan}\ }\textbf {\bibinfo
  {volume} {51}},\ \bibinfo {pages} {839} (\bibinfo {year} {1982})},\ \Eprint
  {https://arxiv.org/abs/https://doi.org/10.1143/JPSJ.51.839}
  {https://doi.org/10.1143/JPSJ.51.839} \BibitemShut {NoStop}%
\bibitem [{\citenamefont {Press}\ \emph {et~al.}(2008)\citenamefont {Press},
  \citenamefont {Ladd}, \citenamefont {Zhang},\ and\ \citenamefont
  {Yamamoto}}]{Press2008}%
  \BibitemOpen
  \bibfield  {author} {\bibinfo {author} {\bibfnamefont {D.}~\bibnamefont
  {Press}}, \bibinfo {author} {\bibfnamefont {T.~D.}\ \bibnamefont {Ladd}},
  \bibinfo {author} {\bibfnamefont {B.}~\bibnamefont {Zhang}},\ and\ \bibinfo
  {author} {\bibfnamefont {Y.}~\bibnamefont {Yamamoto}},\ }\bibfield  {title}
  {\bibinfo {title} {{Complete quantum control of a single quantum dot spin
  using ultrafast optical pulses}},\ }\href
  {https://doi.org/10.1038/nature07530} {\bibfield  {journal} {\bibinfo
  {journal} {Nature}\ }\textbf {\bibinfo {volume} {456}},\ \bibinfo {pages}
  {218} (\bibinfo {year} {2008})}\BibitemShut {NoStop}%
\bibitem [{\citenamefont {Yugova}\ \emph {et~al.}(2007)\citenamefont {Yugova},
  \citenamefont {Greilich}, \citenamefont {Yakovlev}, \citenamefont {Kiselev},
  \citenamefont {Bayer}, \citenamefont {Petrov}, \citenamefont {Dolgikh},
  \citenamefont {Reuter},\ and\ \citenamefont {Wieck}}]{Yugova2007}%
  \BibitemOpen
  \bibfield  {author} {\bibinfo {author} {\bibfnamefont {I.~A.}\ \bibnamefont
  {Yugova}}, \bibinfo {author} {\bibfnamefont {A.}~\bibnamefont {Greilich}},
  \bibinfo {author} {\bibfnamefont {D.~R.}\ \bibnamefont {Yakovlev}}, \bibinfo
  {author} {\bibfnamefont {A.~A.}\ \bibnamefont {Kiselev}}, \bibinfo {author}
  {\bibfnamefont {M.}~\bibnamefont {Bayer}}, \bibinfo {author} {\bibfnamefont
  {V.~V.}\ \bibnamefont {Petrov}}, \bibinfo {author} {\bibfnamefont {Y.~K.}\
  \bibnamefont {Dolgikh}}, \bibinfo {author} {\bibfnamefont {D.}~\bibnamefont
  {Reuter}},\ and\ \bibinfo {author} {\bibfnamefont {A.~D.}\ \bibnamefont
  {Wieck}},\ }\bibfield  {title} {\bibinfo {title} {Universal behavior of the
  electron $g$ factor in
  $\mathrm{Ga}\mathrm{As}/{\mathrm{al}}_{x}{\mathrm{ga}}_{1\ensuremath{-}x}\mathrm{As}$
  quantum wells},\ }\href {https://doi.org/10.1103/PhysRevB.75.245302}
  {\bibfield  {journal} {\bibinfo  {journal} {Phys. Rev. B}\ }\textbf {\bibinfo
  {volume} {75}},\ \bibinfo {pages} {245302} (\bibinfo {year}
  {2007})}\BibitemShut {NoStop}%
\bibitem [{\citenamefont {Yang}\ \emph {et~al.}(2010)\citenamefont {Yang},
  \citenamefont {Dai}, \citenamefont {Ge},\ and\ \citenamefont
  {Cui}}]{Yang2010}%
  \BibitemOpen
  \bibfield  {author} {\bibinfo {author} {\bibfnamefont {C.~L.}\ \bibnamefont
  {Yang}}, \bibinfo {author} {\bibfnamefont {J.}~\bibnamefont {Dai}}, \bibinfo
  {author} {\bibfnamefont {W.~K.}\ \bibnamefont {Ge}},\ and\ \bibinfo {author}
  {\bibfnamefont {X.}~\bibnamefont {Cui}},\ }\bibfield  {title} {\bibinfo
  {title} {{Determination of the sign of g factors for conduction electrons
  using time-resolved Kerr rotation}},\ }\href
  {https://doi.org/10.1063/1.3402769} {\bibfield  {journal} {\bibinfo
  {journal} {Applied Physics Letters}\ }\textbf {\bibinfo {volume} {96}},\
  \bibinfo {pages} {1} (\bibinfo {year} {2010})}\BibitemShut {NoStop}%
\bibitem [{\citenamefont {Bu{\ss}}\ \emph {et~al.}(2015)\citenamefont
  {Bu{\ss}}, \citenamefont {Schupp}, \citenamefont {As}, \citenamefont
  {H{\"{a}}gele},\ and\ \citenamefont {Rudolph}}]{Rudolph2015}%
  \BibitemOpen
  \bibfield  {author} {\bibinfo {author} {\bibfnamefont {J.~H.}\ \bibnamefont
  {Bu{\ss}}}, \bibinfo {author} {\bibfnamefont {T.}~\bibnamefont {Schupp}},
  \bibinfo {author} {\bibfnamefont {D.~J.}\ \bibnamefont {As}}, \bibinfo
  {author} {\bibfnamefont {D.}~\bibnamefont {H{\"{a}}gele}},\ and\ \bibinfo
  {author} {\bibfnamefont {J.}~\bibnamefont {Rudolph}},\ }\bibfield  {title}
  {\bibinfo {title} {{Temperature dependence of the electron Land{\'{e}} g
  -factor in cubic GaN}},\ }\href {https://doi.org/10.1063/1.4937128}
  {\bibfield  {journal} {\bibinfo  {journal} {Journal of Applied Physics}\
  }\textbf {\bibinfo {volume} {118}},\ \bibinfo {pages} {225701} (\bibinfo
  {year} {2015})}\BibitemShut {NoStop}%
\bibitem [{\citenamefont {Yang}\ \emph {et~al.}(2015)\citenamefont {Yang},
  \citenamefont {Chen}, \citenamefont {McCreary}, \citenamefont {Jonker},
  \citenamefont {Lou},\ and\ \citenamefont {Crooker}}]{Yang2015}%
  \BibitemOpen
  \bibfield  {author} {\bibinfo {author} {\bibfnamefont {L.}~\bibnamefont
  {Yang}}, \bibinfo {author} {\bibfnamefont {W.}~\bibnamefont {Chen}}, \bibinfo
  {author} {\bibfnamefont {K.~M.}\ \bibnamefont {McCreary}}, \bibinfo {author}
  {\bibfnamefont {B.~T.}\ \bibnamefont {Jonker}}, \bibinfo {author}
  {\bibfnamefont {J.}~\bibnamefont {Lou}},\ and\ \bibinfo {author}
  {\bibfnamefont {S.~A.}\ \bibnamefont {Crooker}},\ }\bibfield  {title}
  {\bibinfo {title} {{Spin Coherence and Dephasing of Localized Electrons in
  Monolayer MoS$_2$}},\ }\href {https://doi.org/10.1021/acs.nanolett.5b03771}
  {\bibfield  {journal} {\bibinfo  {journal} {Nano Letters}\ }\textbf {\bibinfo
  {volume} {15}},\ \bibinfo {pages} {8250} (\bibinfo {year} {2015})},\ \Eprint
  {https://arxiv.org/abs/1511.03255} {arXiv:1511.03255} \BibitemShut {NoStop}%
\bibitem [{\citenamefont {McCormick}\ \emph {et~al.}(2017)\citenamefont
  {McCormick}, \citenamefont {Newburger}, \citenamefont {Luo}, \citenamefont
  {McCreary}, \citenamefont {Singh}, \citenamefont {Martin}, \citenamefont
  {Cichewicz}, \citenamefont {Jonker},\ and\ \citenamefont
  {Kawakami}}]{McCormick2017}%
  \BibitemOpen
  \bibfield  {author} {\bibinfo {author} {\bibfnamefont {E.~J.}\ \bibnamefont
  {McCormick}}, \bibinfo {author} {\bibfnamefont {M.~J.}\ \bibnamefont
  {Newburger}}, \bibinfo {author} {\bibfnamefont {Y.~K.}\ \bibnamefont {Luo}},
  \bibinfo {author} {\bibfnamefont {K.~M.}\ \bibnamefont {McCreary}}, \bibinfo
  {author} {\bibfnamefont {S.}~\bibnamefont {Singh}}, \bibinfo {author}
  {\bibfnamefont {I.~B.}\ \bibnamefont {Martin}}, \bibinfo {author}
  {\bibfnamefont {E.~J.}\ \bibnamefont {Cichewicz}}, \bibinfo {author}
  {\bibfnamefont {B.~T.}\ \bibnamefont {Jonker}},\ and\ \bibinfo {author}
  {\bibfnamefont {R.~K.}\ \bibnamefont {Kawakami}},\ }\bibfield  {title}
  {\bibinfo {title} {{Imaging spin dynamics in monolayer WS$_2$ by
  time-resolved Kerr rotation microscopy}},\ }\href
  {https://doi.org/10.1088/2053-1583/aa98ae} {\bibfield  {journal} {\bibinfo
  {journal} {2D Materials}\ }\textbf {\bibinfo {volume} {5}},\ \bibinfo {pages}
  {011010} (\bibinfo {year} {2017})},\ \Eprint
  {https://arxiv.org/abs/1602.03568} {arXiv:1602.03568} \BibitemShut {NoStop}%
\bibitem [{\citenamefont {Zhu}\ \emph {et~al.}(2014)\citenamefont {Zhu},
  \citenamefont {Zhang}, \citenamefont {Glazov}, \citenamefont {Urbaszek},
  \citenamefont {Amand}, \citenamefont {Ji}, \citenamefont {Liu},\ and\
  \citenamefont {Marie}}]{Zhu2014}%
  \BibitemOpen
  \bibfield  {author} {\bibinfo {author} {\bibfnamefont {C.~R.}\ \bibnamefont
  {Zhu}}, \bibinfo {author} {\bibfnamefont {K.}~\bibnamefont {Zhang}}, \bibinfo
  {author} {\bibfnamefont {M.}~\bibnamefont {Glazov}}, \bibinfo {author}
  {\bibfnamefont {B.}~\bibnamefont {Urbaszek}}, \bibinfo {author}
  {\bibfnamefont {T.}~\bibnamefont {Amand}}, \bibinfo {author} {\bibfnamefont
  {Z.~W.}\ \bibnamefont {Ji}}, \bibinfo {author} {\bibfnamefont {B.~L.}\
  \bibnamefont {Liu}},\ and\ \bibinfo {author} {\bibfnamefont {X.}~\bibnamefont
  {Marie}},\ }\bibfield  {title} {\bibinfo {title} {{Exciton valley dynamics
  probed by Kerr rotation in ${\mathrm{WSe}}_{2}$ monolayers}},\ }\href
  {https://doi.org/10.1103/PhysRevB.90.161302} {\bibfield  {journal} {\bibinfo
  {journal} {Phys. Rev. B}\ }\textbf {\bibinfo {volume} {90}},\ \bibinfo
  {pages} {161302(R)} (\bibinfo {year} {2014})}\BibitemShut {NoStop}%
\bibitem [{\citenamefont {LaMountain}\ \emph {et~al.}(2018)\citenamefont
  {LaMountain}, \citenamefont {Bergeron}, \citenamefont {Balla}, \citenamefont
  {Stanev}, \citenamefont {Hersam},\ and\ \citenamefont
  {Stern}}]{LaMountain2018}%
  \BibitemOpen
  \bibfield  {author} {\bibinfo {author} {\bibfnamefont {T.}~\bibnamefont
  {LaMountain}}, \bibinfo {author} {\bibfnamefont {H.}~\bibnamefont
  {Bergeron}}, \bibinfo {author} {\bibfnamefont {I.}~\bibnamefont {Balla}},
  \bibinfo {author} {\bibfnamefont {T.~K.}\ \bibnamefont {Stanev}}, \bibinfo
  {author} {\bibfnamefont {M.~C.}\ \bibnamefont {Hersam}},\ and\ \bibinfo
  {author} {\bibfnamefont {N.~P.}\ \bibnamefont {Stern}},\ }\bibfield  {title}
  {\bibinfo {title} {{Valley-selective optical Stark effect probed by Kerr
  rotation}},\ }\href {https://doi.org/10.1103/PhysRevB.97.045307} {\bibfield
  {journal} {\bibinfo  {journal} {Phys. Rev. B}\ }\textbf {\bibinfo {volume}
  {97}},\ \bibinfo {pages} {045307} (\bibinfo {year} {2018})}\BibitemShut
  {NoStop}%
\bibitem [{\citenamefont {Rybkovskiy}\ \emph {et~al.}(2014)\citenamefont
  {Rybkovskiy}, \citenamefont {Osadchy},\ and\ \citenamefont
  {Obraztsova}}]{Rybkovskiy2014}%
  \BibitemOpen
  \bibfield  {author} {\bibinfo {author} {\bibfnamefont {D.~V.}\ \bibnamefont
  {Rybkovskiy}}, \bibinfo {author} {\bibfnamefont {A.~V.}\ \bibnamefont
  {Osadchy}},\ and\ \bibinfo {author} {\bibfnamefont {E.~D.}\ \bibnamefont
  {Obraztsova}},\ }\bibfield  {title} {\bibinfo {title} {{Transition from
  parabolic to ring-shaped valence band maximum in few-layer GaS, GaSe, and
  InSe}},\ }\href {https://doi.org/10.1103/PhysRevB.90.235302} {\bibfield
  {journal} {\bibinfo  {journal} {Phys. Rev. B}\ }\textbf {\bibinfo {volume}
  {90}},\ \bibinfo {pages} {235302} (\bibinfo {year} {2014})}\BibitemShut
  {NoStop}%
\bibitem [{\citenamefont {Takasuna}\ \emph {et~al.}(2017)\citenamefont
  {Takasuna}, \citenamefont {Shiogai}, \citenamefont {Matsuzaka}, \citenamefont
  {Kohda}, \citenamefont {Oyama},\ and\ \citenamefont {Nitta}}]{Takasuna2017}%
  \BibitemOpen
  \bibfield  {author} {\bibinfo {author} {\bibfnamefont {S.}~\bibnamefont
  {Takasuna}}, \bibinfo {author} {\bibfnamefont {J.}~\bibnamefont {Shiogai}},
  \bibinfo {author} {\bibfnamefont {S.}~\bibnamefont {Matsuzaka}}, \bibinfo
  {author} {\bibfnamefont {M.}~\bibnamefont {Kohda}}, \bibinfo {author}
  {\bibfnamefont {Y.}~\bibnamefont {Oyama}},\ and\ \bibinfo {author}
  {\bibfnamefont {J.}~\bibnamefont {Nitta}},\ }\bibfield  {title} {\bibinfo
  {title} {{Weak antilocalization induced by Rashba spin-orbit interaction in
  layered III-VI compound semiconductor GaSe thin films}},\ }\href
  {https://doi.org/10.1103/PhysRevB.96.161303} {\bibfield  {journal} {\bibinfo
  {journal} {Phys. Rev. B}\ }\textbf {\bibinfo {volume} {96}},\ \bibinfo
  {pages} {161303(R)} (\bibinfo {year} {2017})}\BibitemShut {NoStop}%
\bibitem [{\citenamefont {Wu}\ \emph {et~al.}(2010)\citenamefont {Wu},
  \citenamefont {Jiang},\ and\ \citenamefont {Weng}}]{Wu2010}%
  \BibitemOpen
  \bibfield  {author} {\bibinfo {author} {\bibfnamefont {M.~W.}\ \bibnamefont
  {Wu}}, \bibinfo {author} {\bibfnamefont {J.~H.}\ \bibnamefont {Jiang}},\ and\
  \bibinfo {author} {\bibfnamefont {M.~Q.}\ \bibnamefont {Weng}},\ }\bibfield
  {title} {\bibinfo {title} {{Spin dynamics in semiconductors}},\ }\href
  {https://doi.org/10.1016/j.physrep.2010.04.002} {\bibfield  {journal}
  {\bibinfo  {journal} {Physics Reports}\ }\textbf {\bibinfo {volume} {493}},\
  \bibinfo {pages} {61} (\bibinfo {year} {2010})},\ \Eprint
  {https://arxiv.org/abs/1001.0606} {arXiv:1001.0606} \BibitemShut {NoStop}%
\bibitem [{\citenamefont {Hermann}\ and\ \citenamefont
  {Weisbuch}(1977)}]{Hermann1977}%
  \BibitemOpen
  \bibfield  {author} {\bibinfo {author} {\bibfnamefont {C.}~\bibnamefont
  {Hermann}}\ and\ \bibinfo {author} {\bibfnamefont {C.}~\bibnamefont
  {Weisbuch}},\ }\bibfield  {title} {\bibinfo {title} {{k $\cdot$ p perturbation
  theory in III-V compounds and alloys: a reexamination}},\ }\href
  {https://doi.org/10.1103/PhysRevB.15.823} {\bibfield  {journal} {\bibinfo
  {journal} {Physical Review B}\ }\textbf {\bibinfo {volume} {15}},\ \bibinfo
  {pages} {823} (\bibinfo {year} {1977})}\BibitemShut {NoStop}%
\bibitem [{\citenamefont {Wang}\ \emph {et~al.}(2015)\citenamefont {Wang},
  \citenamefont {Bouet}, \citenamefont {Glazov}, \citenamefont {Amand},
  \citenamefont {Ivchenko}, \citenamefont {Palleau}, \citenamefont {Marie},\
  and\ \citenamefont {Urbaszek}}]{Wang2015}%
  \BibitemOpen
  \bibfield  {author} {\bibinfo {author} {\bibfnamefont {G.}~\bibnamefont
  {Wang}}, \bibinfo {author} {\bibfnamefont {L.}~\bibnamefont {Bouet}},
  \bibinfo {author} {\bibfnamefont {M.~M.}\ \bibnamefont {Glazov}}, \bibinfo
  {author} {\bibfnamefont {T.}~\bibnamefont {Amand}}, \bibinfo {author}
  {\bibfnamefont {E.~L.}\ \bibnamefont {Ivchenko}}, \bibinfo {author}
  {\bibfnamefont {E.}~\bibnamefont {Palleau}}, \bibinfo {author} {\bibfnamefont
  {X.}~\bibnamefont {Marie}},\ and\ \bibinfo {author} {\bibfnamefont
  {B.}~\bibnamefont {Urbaszek}},\ }\bibfield  {title} {\bibinfo {title}
  {{Magneto-optics in transition metal diselenide monolayers}},\ }\bibfield
  {journal} {\bibinfo  {journal} {2D Materials}\ }\textbf {\bibinfo {volume}
  {2}},\ \href {https://doi.org/10.1088/2053-1583/2/3/034002}
  {10.1088/2053-1583/2/3/034002} (\bibinfo {year} {2015}),\ \Eprint
  {https://arxiv.org/abs/1503.04105} {arXiv:1503.04105} \BibitemShut {NoStop}%
\bibitem [{\citenamefont {Yao}\ and\ \citenamefont {Zhang}(2021)}]{yao21}%
  \BibitemOpen
  \bibfield  {author} {\bibinfo {author} {\bibfnamefont {X.}~\bibnamefont
  {Yao}}\ and\ \bibinfo {author} {\bibfnamefont {X.}~\bibnamefont {Zhang}},\
  }\bibfield  {title} {\bibinfo {title} {{Electronic Structures of Twisted
  Bilayer InSe/InSe and Heterobilayer Graphene/InSe}},\ }\href@noop {}
  {\bibfield  {journal} {\bibinfo  {journal} {ACS\ Omega}\ }\textbf {\bibinfo
  {volume} {6}},\ \bibinfo {pages} {13426} (\bibinfo {year}
  {2021})}\BibitemShut {NoStop}%
\bibitem [{\citenamefont {F$\text{\"{o}}$rste}\ \emph
  {et~al.}(2020)\citenamefont {F$\text{\"{o}}$rste}, \citenamefont {Tepliakov},
  \citenamefont {Kruchinin}, \citenamefont {Lindlau}, \citenamefont {Funk},
  \citenamefont {F$\text{\"{o}}$rg}, \citenamefont {Watanabe}, \citenamefont
  {Taniguchi}, \citenamefont {Baimuratov},\ and\ \citenamefont
  {H$\text{\"{o}}$gele}}]{forste20}%
  \BibitemOpen
  \bibfield  {author} {\bibinfo {author} {\bibfnamefont {J.}~\bibnamefont
  {F$\text{\"{o}}$rste}}, \bibinfo {author} {\bibfnamefont {N.~V.}\
  \bibnamefont {Tepliakov}}, \bibinfo {author} {\bibfnamefont {S.~Y.}\
  \bibnamefont {Kruchinin}}, \bibinfo {author} {\bibfnamefont {J.}~\bibnamefont
  {Lindlau}}, \bibinfo {author} {\bibfnamefont {V.}~\bibnamefont {Funk}},
  \bibinfo {author} {\bibfnamefont {M.}~\bibnamefont {F$\text{\"{o}}$rg}},
  \bibinfo {author} {\bibfnamefont {K.}~\bibnamefont {Watanabe}}, \bibinfo
  {author} {\bibfnamefont {T.}~\bibnamefont {Taniguchi}}, \bibinfo {author}
  {\bibfnamefont {A.~S.}\ \bibnamefont {Baimuratov}},\ and\ \bibinfo {author}
  {\bibfnamefont {A.}~\bibnamefont {H$\text{\"{o}}$gele}},\ }\bibfield  {title}
  {\bibinfo {title} {{Exciton g-factors in monolayer and bilayer WSe$_2$ from
  experiment and theory}},\ }\href@noop {} {\bibfield  {journal} {\bibinfo
  {journal} {Nat.\ Commun.}\ }\textbf {\bibinfo {volume} {112}},\ \bibinfo
  {pages} {4539} (\bibinfo {year} {2020})}\BibitemShut {NoStop}%
\bibitem [{\citenamefont {F$\text{\"{o}}$rg}\ \emph {et~al.}(2021)\citenamefont
  {F$\text{\"{o}}$rg}, \citenamefont {Baimuratov}, \citenamefont {Kruchinin},
  \citenamefont {Vovk}, \citenamefont {Scherzer}, \citenamefont
  {F$\text{\"{o}}$rste}, \citenamefont {Funk}, \citenamefont {Watanabe},
  \citenamefont {Taniguchi},\ and\ \citenamefont
  {H$\text{\"{o}}$gele}}]{forg21}%
  \BibitemOpen
  \bibfield  {author} {\bibinfo {author} {\bibfnamefont {M.}~\bibnamefont
  {F$\text{\"{o}}$rg}}, \bibinfo {author} {\bibfnamefont {A.~S.}\ \bibnamefont
  {Baimuratov}}, \bibinfo {author} {\bibfnamefont {S.~Y.}\ \bibnamefont
  {Kruchinin}}, \bibinfo {author} {\bibfnamefont {I.~A.}\ \bibnamefont {Vovk}},
  \bibinfo {author} {\bibfnamefont {J.}~\bibnamefont {Scherzer}}, \bibinfo
  {author} {\bibfnamefont {J.}~\bibnamefont {F$\text{\"{o}}$rste}}, \bibinfo
  {author} {\bibfnamefont {V.}~\bibnamefont {Funk}}, \bibinfo {author}
  {\bibfnamefont {K.}~\bibnamefont {Watanabe}}, \bibinfo {author}
  {\bibfnamefont {T.}~\bibnamefont {Taniguchi}},\ and\ \bibinfo {author}
  {\bibfnamefont {A.}~\bibnamefont {H$\text{\"{o}}$gele}},\ }\bibfield  {title}
  {\bibinfo {title} {{Moiré excitons in MoSe$_2$-WSe$_2$ heterobilayers and
  heterotrilayers}},\ }\href@noop {} {\bibfield  {journal} {\bibinfo  {journal}
  {Nat.\ Commun.}\ }\textbf {\bibinfo {volume} {12}},\ \bibinfo {pages} {1656}
  (\bibinfo {year} {2021})}\BibitemShut {NoStop}%
\bibitem [{\citenamefont {Roth}\ \emph {et~al.}(1959)\citenamefont {Roth},
  \citenamefont {Lax},\ and\ \citenamefont {Zwerdling}}]{roth59}%
  \BibitemOpen
  \bibfield  {author} {\bibinfo {author} {\bibfnamefont {L.~M.}\ \bibnamefont
  {Roth}}, \bibinfo {author} {\bibfnamefont {B.}~\bibnamefont {Lax}},\ and\
  \bibinfo {author} {\bibfnamefont {S.}~\bibnamefont {Zwerdling}},\ }\bibfield
  {title} {\bibinfo {title} {Theory of optical magneto-absorption effects in
  semiconductors},\ }\href@noop {} {\bibfield  {journal} {\bibinfo  {journal}
  {Phys.\ Rev.}\ }\textbf {\bibinfo {volume} {114}},\ \bibinfo {pages} {90}
  (\bibinfo {year} {1959})}\BibitemShut {NoStop}%
\bibitem [{\citenamefont {Xiao}\ \emph {et~al.}(2010)\citenamefont {Xiao},
  \citenamefont {Chang},\ and\ \citenamefont {Niu}}]{xiao10}%
  \BibitemOpen
  \bibfield  {author} {\bibinfo {author} {\bibfnamefont {D.}~\bibnamefont
  {Xiao}}, \bibinfo {author} {\bibfnamefont {M.-C.}\ \bibnamefont {Chang}},\
  and\ \bibinfo {author} {\bibfnamefont {Q.}~\bibnamefont {Niu}},\ }\bibfield
  {title} {\bibinfo {title} {Berry phase effects on electronic properties},\
  }\href@noop {} {\bibfield  {journal} {\bibinfo  {journal} {Rev.\ Mod.\
  Phys.}\ }\textbf {\bibinfo {volume} {82}},\ \bibinfo {pages} {1959} (\bibinfo
  {year} {2010})}\BibitemShut {NoStop}%
\bibitem [{\citenamefont {Perdew}\ \emph {et~al.}(1996)\citenamefont {Perdew},
  \citenamefont {Burke},\ and\ \citenamefont {Ernzerhof}}]{perdew97}%
  \BibitemOpen
  \bibfield  {author} {\bibinfo {author} {\bibfnamefont {J.~P.}\ \bibnamefont
  {Perdew}}, \bibinfo {author} {\bibfnamefont {K.}~\bibnamefont {Burke}},\ and\
  \bibinfo {author} {\bibfnamefont {M.}~\bibnamefont {Ernzerhof}},\ }\bibfield
  {title} {\bibinfo {title} {{Generalized Gradient Approximation Made
  Simple}},\ }\href {https://doi.org/10.1103/PhysRevLett.77.3865} {\bibfield
  {journal} {\bibinfo  {journal} {Phys. Rev. Lett.}\ }\textbf {\bibinfo
  {volume} {77}},\ \bibinfo {pages} {3865} (\bibinfo {year}
  {1996})}\BibitemShut {NoStop}%
\bibitem [{\citenamefont {Giannozzi}\ \emph {et~al.}(2009)\citenamefont
  {Giannozzi}, \citenamefont {Baroni}, \citenamefont {Bonini}, \citenamefont
  {Calandra}, \citenamefont {Car}, \citenamefont {Cavazzoni}, \citenamefont
  {Ceresoli}, \citenamefont {Chiarotti}, \citenamefont {Cococcioni},
  \citenamefont {Dabo}, \citenamefont {Corso}, \citenamefont {de~Gironcoli},
  \citenamefont {Fabris}, \citenamefont {Fratesi}, \citenamefont {Gebauer},
  \citenamefont {Gerstmann}, \citenamefont {Gougoussis}, \citenamefont
  {Kokalj}, \citenamefont {Lazzeri}, \citenamefont {M-.Samos}, \citenamefont
  {Mazari}, \citenamefont {Mauri}, \citenamefont {Mazzarello}, \citenamefont
  {Paolini}, \citenamefont {Pasquarello}, \citenamefont {Paulatto},
  \citenamefont {Sbraccia}, \citenamefont {Scandolo}, \citenamefont
  {Sclauzero}, \citenamefont {Seitsonen}, \citenamefont {Smogunov},
  \citenamefont {Umari},\ and\ \citenamefont {Wentzcovitch}}]{giannozzi09}%
  \BibitemOpen
  \bibfield  {author} {\bibinfo {author} {\bibfnamefont {P.}~\bibnamefont
  {Giannozzi}}, \bibinfo {author} {\bibfnamefont {S.}~\bibnamefont {Baroni}},
  \bibinfo {author} {\bibfnamefont {N.}~\bibnamefont {Bonini}}, \bibinfo
  {author} {\bibfnamefont {M.}~\bibnamefont {Calandra}}, \bibinfo {author}
  {\bibfnamefont {R.}~\bibnamefont {Car}}, \bibinfo {author} {\bibfnamefont
  {C.}~\bibnamefont {Cavazzoni}}, \bibinfo {author} {\bibfnamefont
  {D.}~\bibnamefont {Ceresoli}}, \bibinfo {author} {\bibfnamefont {G.~L.}\
  \bibnamefont {Chiarotti}}, \bibinfo {author} {\bibfnamefont {M.}~\bibnamefont
  {Cococcioni}}, \bibinfo {author} {\bibfnamefont {I.}~\bibnamefont {Dabo}},
  \bibinfo {author} {\bibfnamefont {A.~D.}\ \bibnamefont {Corso}}, \bibinfo
  {author} {\bibfnamefont {S.}~\bibnamefont {de~Gironcoli}}, \bibinfo {author}
  {\bibfnamefont {S.}~\bibnamefont {Fabris}}, \bibinfo {author} {\bibfnamefont
  {G.}~\bibnamefont {Fratesi}}, \bibinfo {author} {\bibfnamefont
  {R.}~\bibnamefont {Gebauer}}, \bibinfo {author} {\bibfnamefont
  {U.}~\bibnamefont {Gerstmann}}, \bibinfo {author} {\bibfnamefont
  {C.}~\bibnamefont {Gougoussis}}, \bibinfo {author} {\bibfnamefont
  {A.}~\bibnamefont {Kokalj}}, \bibinfo {author} {\bibfnamefont
  {M.}~\bibnamefont {Lazzeri}}, \bibinfo {author} {\bibfnamefont
  {L.}~\bibnamefont {M-.Samos}}, \bibinfo {author} {\bibfnamefont
  {N.}~\bibnamefont {Mazari}}, \bibinfo {author} {\bibfnamefont
  {F.}~\bibnamefont {Mauri}}, \bibinfo {author} {\bibfnamefont
  {R.}~\bibnamefont {Mazzarello}}, \bibinfo {author} {\bibfnamefont
  {S.}~\bibnamefont {Paolini}}, \bibinfo {author} {\bibfnamefont
  {A.}~\bibnamefont {Pasquarello}}, \bibinfo {author} {\bibfnamefont
  {L.}~\bibnamefont {Paulatto}}, \bibinfo {author} {\bibfnamefont
  {C.}~\bibnamefont {Sbraccia}}, \bibinfo {author} {\bibfnamefont
  {S.}~\bibnamefont {Scandolo}}, \bibinfo {author} {\bibfnamefont
  {G.}~\bibnamefont {Sclauzero}}, \bibinfo {author} {\bibfnamefont {A.~P.}\
  \bibnamefont {Seitsonen}}, \bibinfo {author} {\bibfnamefont {A.}~\bibnamefont
  {Smogunov}}, \bibinfo {author} {\bibfnamefont {P.}~\bibnamefont {Umari}},\
  and\ \bibinfo {author} {\bibfnamefont {R.~M.}\ \bibnamefont {Wentzcovitch}},\
  }\bibfield  {title} {\bibinfo {title} {{QUANTUM ESPRESSO: a modular and
  open-source software project for quantum simulations of materials}},\
  }\href@noop {} {\bibfield  {journal} {\bibinfo  {journal} {J.\ Phys.:\
  Condens.\ Matter}\ }\textbf {\bibinfo {volume} {21}},\ \bibinfo {pages}
  {395502} (\bibinfo {year} {2009})}\BibitemShut {NoStop}%
\bibitem [{\citenamefont {Burkatzki}\ \emph {et~al.}(2007)\citenamefont
  {Burkatzki}, \citenamefont {Filippi},\ and\ \citenamefont
  {Dolg}}]{burkatzki07}%
  \BibitemOpen
  \bibfield  {author} {\bibinfo {author} {\bibfnamefont {M.}~\bibnamefont
  {Burkatzki}}, \bibinfo {author} {\bibfnamefont {C.}~\bibnamefont {Filippi}},\
  and\ \bibinfo {author} {\bibfnamefont {M.}~\bibnamefont {Dolg}},\ }\bibfield
  {title} {\bibinfo {title} {{Energy-consistent pseudopotentials for quantum
  Monte Carlo calculations}},\ }\href@noop {} {\bibfield  {journal} {\bibinfo
  {journal} {J.\ Chem.\ Phys.}\ }\textbf {\bibinfo {volume} {126}},\ \bibinfo
  {pages} {234105} (\bibinfo {year} {2007})}\BibitemShut {NoStop}%
\bibitem [{\citenamefont {Burkatzki}\ \emph {et~al.}(2008)\citenamefont
  {Burkatzki}, \citenamefont {Filippi},\ and\ \citenamefont
  {Dolg}}]{burkatzki08}%
  \BibitemOpen
  \bibfield  {author} {\bibinfo {author} {\bibfnamefont {M.}~\bibnamefont
  {Burkatzki}}, \bibinfo {author} {\bibfnamefont {C.}~\bibnamefont {Filippi}},\
  and\ \bibinfo {author} {\bibfnamefont {M.}~\bibnamefont {Dolg}},\ }\bibfield
  {title} {\bibinfo {title} {{Energy-consistent small-core pseudopotentials for
  3d-transition metals adapted to quantum Monte Carlo calculations}},\
  }\href@noop {} {\bibfield  {journal} {\bibinfo  {journal} {J.\ Chem.\ Phys.}\
  }\textbf {\bibinfo {volume} {129}},\ \bibinfo {pages} {164115} (\bibinfo
  {year} {2008})}\BibitemShut {NoStop}%
\bibitem [{\citenamefont {Grimme}\ \emph {et~al.}(2010)\citenamefont {Grimme},
  \citenamefont {Antony}, \citenamefont {Ehrlich},\ and\ \citenamefont
  {Krieg}}]{grimme10}%
  \BibitemOpen
  \bibfield  {author} {\bibinfo {author} {\bibfnamefont {S.}~\bibnamefont
  {Grimme}}, \bibinfo {author} {\bibfnamefont {J.}~\bibnamefont {Antony}},
  \bibinfo {author} {\bibfnamefont {S.}~\bibnamefont {Ehrlich}},\ and\ \bibinfo
  {author} {\bibfnamefont {H.}~\bibnamefont {Krieg}},\ }\bibfield  {title}
  {\bibinfo {title} {{A consistent and accurate ab initio parametrization of
  density functional dispersion correction (DFT-D) for the 94 elements H-Pu}},\
  }\href@noop {} {\bibfield  {journal} {\bibinfo  {journal} {J.\ Chem.\ Phys.}\
  }\textbf {\bibinfo {volume} {132}},\ \bibinfo {pages} {154104} (\bibinfo
  {year} {2010})}\BibitemShut {NoStop}%
\end{thebibliography}%

\end{document}